\documentclass[twocolumn]{aastex631}
\pdfoutput=1 
\usepackage{amsmath,amstext}
\usepackage[T1]{fontenc}
\usepackage[figure,figure*]{hypcap}
\usepackage{mwe}
\usepackage{lipsum}
\usepackage{graphicx} 
\usepackage{multirow}
\usepackage{xcolor,colortbl}
\usepackage{xspace}
\usepackage{csvsimple}
\usepackage{afterpage}

\newcommand{\twelveCO}{$^{12}$CO}

\newcommand{\XCO}{$X_\text{CO}$\xspace}

\newcommand{\alphavir}{$\alpha_\text{vir}$\xspace}

\shorttitle{ALMA-LEGUS II: The Influence of Sub-Galactic Environment}
\shortauthors{Finn et al.}

\begin{document}

\title{ALMA-LEGUS II: The Influence of Sub-Galactic Environment on Molecular Cloud Properties}

\author[0000-0001-9338-2594]{Molly K. Finn} %
\affiliation{Department of Astronomy, University of Virginia, Charlottesville, VA 22904, USA}

\author[0000-0001-8348-2671]{Kelsey E. Johnson} %
\affiliation{Department of Astronomy, University of Virginia, Charlottesville, VA 22904, USA}

\author[0000-0002-4663-6827]{Remy Indebetouw} %
\affiliation{Department of Astronomy, University of Virginia, Charlottesville, VA 22904, USA}
\affiliation{National Radio Astronomy Observatory, 520 Edgemont Road, Charlottesville, VA 22903, USA}

\author[0000-0002-7408-7589]{Allison H. Costa} %
\affiliation{National Radio Astronomy Observatory, 520 Edgemont Road, Charlottesville, VA 22903, USA}

\author{Angela Adamo}
\affiliation{Department of Astronomy, The Oskar Klein Centre, Stockholm University, SE-106 91 Stockholm, Sweden}

\author[0000-0003-4137-882X]{Alessandra Aloisi}
\affiliation{Space Telescope Science Institute, 3700 San Martin Drive, Baltimore, MD 21218, USA}

\author[0000-0002-7845-8498]{Lauren Bittle}
\affiliation{Independent Researcher}

\author[0000-0002-5189-8004]{Daniela Calzetti}
\affiliation{Department of Astronomy, University of Massachusetts Amherst, 710 North Pleasant Street, Amherst, MA 01003, USA}

\author[0000-0002-5782-9093]{Daniel A. Dale}
\affiliation{Department of Physics \& Astronomy, University of Wyoming, Laramie, WY 82071}

\author[0000-0002-4578-297X]{Clare L. Dobbs}
\affiliation{School of Physics and Astronomy, University of Exeter, Stocker Road, Exeter, EX4 4QL, UK}

\author[0000-0002-3106-7676]{Jennifer Donovan Meyer}
\affiliation{National Radio Astronomy Observatory, 520 Edgemont Road, Charlottesville, VA 22903, USA}

\author[0000-0002-1723-6330]{Bruce G. Elmegreen}
\affiliation{IBM Research Division, T. J. Watson Research Center, 1101 Kitchawan Road, Yorktown Heights, NY 10598, USA}

\author[0000-0002-1392-3520]{Debra M. Elmegreen}
\affiliation{Department of Physics and Astronomy, Vassar College, Poughkeepsie, NY 12604, USA}

\author[0000-0001-6676-3842]{Michele Fumagalli}
\affiliation{Dipartimento di Fisica G. Occhialini, Universit\`a degli Studi di Milano Bicocca, Piazza della Scienza 3, 20126 Milano, Italy}
\affiliation{INAF – Osservatorio Astronomico di Trieste, via G. B. Tiepolo 11, I-34143 Trieste, Italy}

\author[0000-0001-8608-0408]{J. S. Gallagher}
\affiliation{Department of Astronomy, University of Wisconsin-Madison\\ 475 North Charter St., Madison, WI 53706}
\altaffiliation{Department of Physics and Astronomy, Macalester College, 1600 Grand Ave. , St. Paul, MN 55105}

\author[0000-0002-3247-5321]{Kathryn Grasha}
\affiliation{Research School of Astronomy and Astrophysics, Australian National University, Cotter Rd., Weston ACT 2612, Australia} 
\affiliation{ARC Centre of Excellence for Astrophysics in 3D (ASTRO-3D), Canberra ACT 2600, Australia}   
\affiliation{Visiting Fellow, Harvard-Smithsonian Center for Astrophysics, 60 Garden Street, Cambridge, MA 02138, USA}

\author[0000-0002-1891-3794]{Eva K. Grebel}
\affiliation{Astronomisches Rechen-Institut, Zentrum f\"ur Astronomie der Universit\"at Heidelberg, M\"onchhofstr.\ 12--14, 69120 Heidelberg, Germany}

\author[0000-0001-5448-1821]{Robert C. Kennicutt}
\affiliation{Steward Observatory, University of Arizona, Tucson, AZ 85719, USA}
\affiliation{George P. and Cynthia W. Mitchell Institute for Fundamental Physics and Astronomy, Texas A\&M University, College Station, TX 77845, USA}

\author[0000-0003-3893-854X]{Mark R. Krumholz}
\affiliation{Research School of Astronomy and Astrophysics, Australian National University, Cotter Rd., Weston ACT 2612, Australia}
\affiliation{ARC Centre of Excellence for Astrophysics in 3D (ASTRO-3D), Canberra ACT 2600, Australia}

\author[0000-0002-2278-9407]{Janice C. Lee}
\affiliation{Space Telescope Science Institute, 3700 San Martin Drive, Baltimore, MD 21218, USA}

\author[0000-0003-1427-2456]{Matteo Messa}
\affiliation{Observatoire de Gen\'eve, Universit\'e de Gen\`eve, Versoix, Switzerland}
\affiliation{The Oskar Klein Centre, Department of Astronomy, Stockholm University, AlbaNova, SE-10691 Stockholm, Sweden}

\author[0000-0001-7069-4026]{Preethi Nair}
\affiliation{Department of Physics and Astronomy, The University of Alabama, Tuscaloosa, AL 35487, USA}

\author[0000-0003-2954-7643]{Elena Sabbi}
\affiliation{Space Telescope Science Institute, 3700 San Martin Drive, Baltimore, MD 21218, USA}

\author[0000-0002-0806-168X]{Linda J. Smith}
\affiliation{Space Telescope Science Institute, 3700 San Martin Drive, Baltimore, MD 21218, USA}

\author[0000-0002-8528-7340]{David A. Thilker}
\affiliation{Department of Physics and Astronomy, The Johns Hopkins University, Baltimore, MD, 21218 USA}

\author{Bradley C. Whitmore}
\affiliation{Space Telescope Science Institute, 3700 San Martin Drive, Baltimore, MD 21218, USA}

\author[0000-0001-8289-3428]{Aida Wofford}
\affiliation{Instituto de Astronom\'ia, Universidad Nacional Aut\'onoma de M\'exico, Unidad Acad\'emica en Ensenada, Km 103 Carr. Tijuana$-$Ensenada, Ensenada, B.C., C.P. 22860, M\'exico}

\begin{abstract}

We compare the molecular cloud properties in sub-galactic regions of two galaxies, barred spiral NGC~1313, which is forming many massive clusters, and flocculent spiral NGC~7793, which is forming significantly fewer massive clusters despite having a similar star formation rate to NGC~1313. 
We find that there are larger variations in cloud properties between different regions within each galaxy than there are between the galaxies on a global scale, especially for NGC~1313. 
There are higher masses, linewidths, pressures, and virial parameters in the arms of NGC~1313 and center of NGC~7793 than in the interarm and outer regions of the galaxies. 
The massive cluster formation of NGC~1313 may be driven by its greater variation in environments, allowing more clouds with the necessary conditions to arise, although no one parameter seems primarily responsible for the difference in star formation. 
Meanwhile NGC~7793 has clouds that are as massive and have as much kinetic energy as clouds in the arms of NGC~1313, but have densities and pressures more similar to the interarm regions and so are less inclined to collapse and form stars. 
The cloud properties in NGC~1313 and NGC~7793 suggest that spiral arms, bars, interarm regions, and flocculent spirals each represent distinct environments with regard to molecular cloud populations.
{We see surprisingly little difference in surface densities between the regions, suggesting that the differences in surface densities frequently seen between arm and interarm regions of lower-resolution studies are indicative of the sparsity of molecular clouds, rather than differences in their true surface density.}

\end{abstract}


\section{Introduction} \label{sec:intro}

\begin{figure*}
    \centering
    \includegraphics[width=0.93\textwidth]{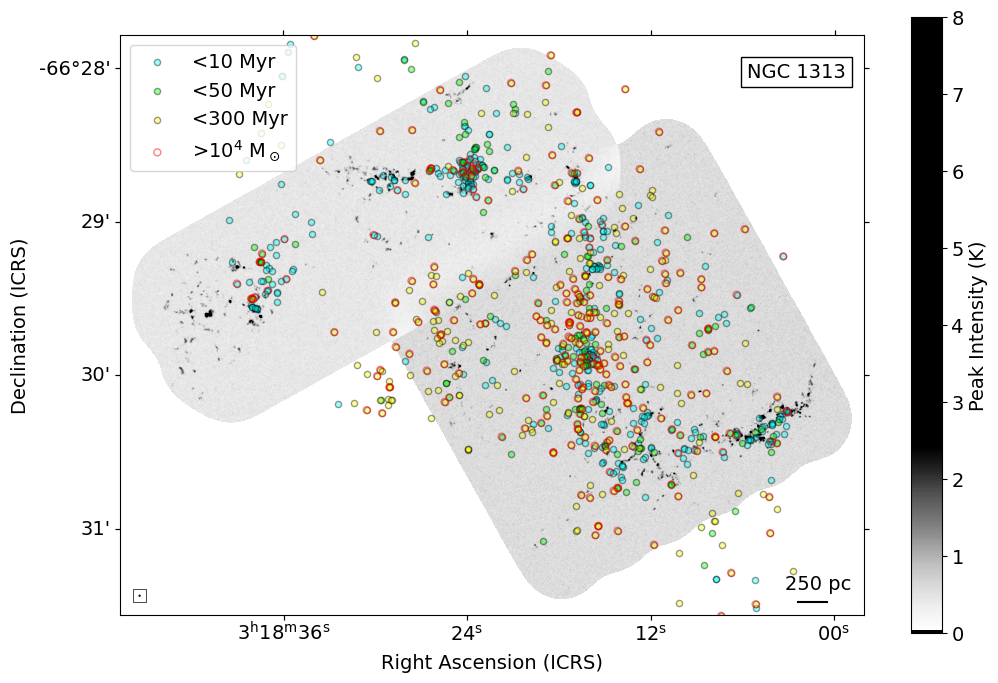}
    \includegraphics[width=0.93\textwidth]{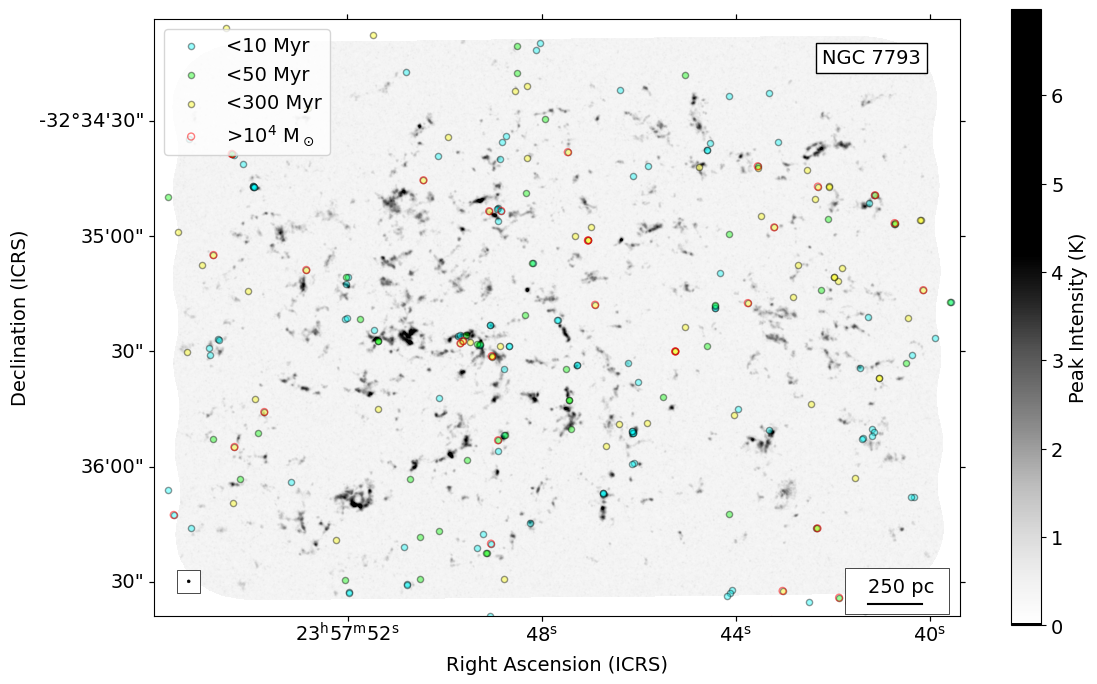}
    \caption{ 
    {Peak intensity maps of CO(2-1) from Paper 1 in grayscale, overlaid with the positions of clusters identified in the LEGUS catalogs for NGC~1313 (top) and NGC~7793 (bottom). The beam sizes of 13~pc are shown in the bottom left corner.} The clusters are colored by age with clusters younger than 10 Myr in cyan, 10-50 Myr in green, and 50-300 Myr in yellow. Clusters that are more massive than $10^4 M_\odot$ are outlined in red. NGC~1313 has significantly more star clusters overall than NGC~7793, and especially has more red-outlined massive star clusters, both by number and by fraction of the total cluster mass. }
    \label{fig:RGB images}
\end{figure*}

Spiral galaxies are home to the majority of local star formation \citep{Brinchmann04}, and so it is important for us to understand how the environments of different regions {within the galaxies} and different types of spiral galaxies affect star formation. The molecular gas in bars, spiral arms, interarm regions, and galaxy centers all experience different conditions, which can in turn influence the star formation taking place in each environment. 

Much work has been done on understanding how spiral density waves and stellar feedback impact cloud formation, collapse, and dispersal, both from simulations and observations. 
There have been several surveys to study the molecular gas in nearby spiral galaxies at the scale of giant molecular clouds (GMCs), such as PAWS (PdBI Arcsecond Whirlpool Survey), which mapped M51 in CO(1-0) at 40~pc resolution \citep{Schinnerer13},
CANON (CArma and NObeyama Nearby galaxies), which mapped the inner disks of nearby spiral galaxies in CO (1-0) and enabled a focused study of molecular cloud properties using a subsample at 62-78 pc resolution \citep{JDM13}, PHANGS-ALMA (Physics at High Angular resolution in Nearby Galaxies), which mapped 90 galaxies in CO(2-1) at $\sim100$~pc resolution \citep{Leroy21}, and most recently the barred spiral galaxy M83 was mapped in CO(1-0) at 40~pc resolution \citep{Koda23}.  

These studies have consistently shown that at $\sim$40-100~pc resolutions, the molecular gas in spiral arms tend to be brighter and have higher surface densities, velocity dispersions, and pressures than gas in the interarm regions, especially when a strong bar is present \citep{Colombo14, Sun18PHANGS, Sun20, Rosolowsky21, Koda23}. They have also shown that the slope of the distribution of cloud masses is shallower and truncates at higher masses in the spiral arms than interarm regions \citep{Koda09, Colombo14, Rosolowsky08}, but that despite the greater amount of star formation in the arms, the depletion time in the arms is not significantly shorter \citep{Querejeta21}. Rather, \cite{Yu21} find that the depletion time of the gas is more closely related to the strength of the spiral arms, with stronger arms being associated with shorter depletion times and higher specific SFR. 

These observations are well-modeled by simulations, especially those of grand design spiral galaxies. Simulations also find that the spiral arms are generally the sites of active star formation rather than the interarm regions \citep{Dobbs13}, and that GMCs are assembled in the spiral arms and are then sheared into smaller clouds in the interarm regions, resulting in lower mass clouds being found in the interarm regions \citep{DobbsPringle13}. Breaking up massive clouds via shear is well-matched to the lifetimes of clouds measured in the interarm region of M51 \citep{Meidt15} and of M83 \citep{Koda23}. Massive clouds are expected to be denser and have longer lifetimes, allowing them to form more massive clusters in the spiral arms before they are dispersed \citep{Dobbs11,Dobbs17}. \cite{Meidt13} propose that the higher amount of streaming motion in the interarm regions relative to the arms stabilizes the clouds and suppresses collapse. 

\cite{Pettitt20} find that the cause of the spiral pattern (e.g. density wave, interaction, or underlying gravitational instability) has no {effect} on the simulated GMC properties. However, GMCs have a shallower mass distribution with more massive clouds after the galaxy experiences a tidal fly-by, especially in the spiral arms, and many of the clouds become unbound during this process \citep{Pettitt18, Nguyen18}. 

In flocculent spiral galaxies, the distinction between interarm and arm regions is less robust, but \cite{Dobbs19} still find a difference in the steepness of the cloud mass distribution, where star-forming clouds have a shallower mass distribution than non-star-forming clouds, similar to the difference between arm and interarm regions in grand design galaxies. However, \cite{Dobbs18} find that to simulate a flocculent spiral with a weak spiral structure, the stellar feedback must be higher than in a galaxy with strong spirals. 

These differences in cloud properties have important repercussions not just for where in the galaxy stars form, but also what kind of stars and star clusters form. Measurements of the cluster formation efficiency, {defined by \cite{Adamo15} as} the fraction of stars that form in clusters, indicate that it is correlated with the surface density of the gas \citep{Adamo11, SilvaVilla13, Adamo15, Johnson16,Adamo20}. 

Furthermore, the maximum mass of star clusters where the mass distribution truncates appears to depend on the SFR surface density of the galaxy \citep{Adamo15,Johnson17,Messa18,Wainer22}. 
A more extreme version of this {trend} has been seen in starburst environments that form massive super star clusters, such as the Antennae galaxies, where clouds have been measured to have extremely high pressures and surface densities \citep{Johnson15, Sun18PHANGS, Finn2019, Krahm23}. The hierarchical clustering of the molecular gas also appears to imprint its structure on the spatial clustering of the star clusters, with implications for the evolution of the star clusters and their potential for dispersal \citep{Grasha17a, Grasha17b, Grasha18, Menon21, Turner22}.

{In this paper, we build upon the analysis of \cite{Finn23a} (hereafter referred to as Paper 1) to further our understanding of how different galactic environments influence the conditions and outcomes of star formation. We use for comparison two galaxies from the Legacy ExtraGalactic UV Survey \citep[LEGUS;][]{Calzetti15}: barred spiral NGC~1313 and flocculent spiral NGC~7793. These galaxies have similar masses \citep[$2.6\times10^9$ and $3.2\times10^9$ M$_\odot$;][]{Calzetti15}, metallicities \citep[$12+\log(\text{O/H})=$ 8.4 and 8.52;][]{WalshRoy97,Stanghellini15}, and star formation rates \citep[SFR; 1.15 and 0.52 M$_\odot$/yr;][]{Calzetti15}. However, in Paper 1 we demonstrate that NGC~1313 hosts significantly more star clusters than NGC~7793 as identified by LEGUS, and especially young, massive clusters, even after correcting for their small difference in star formation rate (see Figure\,\ref{fig:RGB images}). The similarities of the galaxies, combined with their similar distances \citep[4.6 and 3.7 Mpc;][]{RadburnSmith11,Qing15,Gao16,Sabbi18}, makes the pair an excellent laboratory for directly comparing how the molecular gas conditions and galactic structure affect star cluster formation. For a more detailed description of these two galaxies, their properties, and their larger environments, see Paper 1. }

In Paper 1, we found surprisingly minimal differences in the cloud properties when comparing the two galaxies as a whole. NGC~1313 had slightly higher surface densities, external pressures, and lower free-fall times than NGC~7793. The clouds in NGC~1313 also had higher kinetic energies per spatial scale when fitting a power-law to {a} size-linewidth relation, and more clouds were near virial equilibrium than in NGC~7793. However, the clouds with virial parameters near 1 were not any more spatially correlated with the locations of star clusters than the general cloud population in either galaxy. 

{We expand this previous analysis by considering how the sub-galactic environments affect the molecular cloud populations and how those relate to the cluster formation occurring in those regions. We split NGC~1313 into four regions: the bar, northern arm, southern arm, and interarm. In NGC~7793, the spiral arms are too weak for robust delineations of arm and interarm, so we instead split the galaxy into regions by galactocentric radius. These regions are defined in Section\,\ref{sec:obs}, along with the observations, cloud structures, and their properties we use from Paper 1. In Section\,\ref{sec: cluster catalogs} we describe the cluster catalogs used from LEGUS. We examine how the size-linewidth relations of clouds in each region in Section\,\ref{sec:SL plots}, and the virialization of those clouds in Section\,\ref{sec: virial plot}. We then compare the distributions of all the different cloud properties and different galactic regions in Section\,\ref{sec:property comparison}. We discuss our results in Section\,\ref{sec:discussion} and then summarize the primary findings in Section\,\ref{sec:conclusions}. }

\section{Observations} \label{sec:obs}

\subsection{CO(2-1)}

Both galaxies were observed by ALMA (project code 2015.1.00782.S; PI: K. E. Johnson) using Band 6 covering the \twelveCO(2-1) line with the 12m array. The details of these observations, the data reduction, and the imaging process are discussed in Paper 1, {with the parameters of the final images shown in Table\,\ref{tab:observations}}, reproduced from Paper 1. {Figure\,\ref{fig:RGB images} shows the peak intensity maps of these observations, overlaid with the positions of clusters identified by LEGUS.}


\begin{table}
    \centering
    \caption{ALMA \twelveCO(2-1) Observations}
    \begin{tabular}{ccccc}
    \hline
    \hline
        Galaxy & Beam & Beam & rms & Velocity \\
         & (arcsec) & (pc) & (K) & Resolution (km/s)\\
         \hline
        NGC~1313 & 0.58 & 13 & 0.15 & 1.33 \\
        NGC~7793 & 0.72 & 13 & 0.2  & 1.33 \\
        \hline
    \end{tabular}
    
    \label{tab:observations}
\end{table}

In this study, we use the clumps and dendrogram structures identified in Section~4 of Paper 1 using \texttt{quickclump} \citep{Sidorin17} and \texttt{astrodendro} \citep{Rosolowsky08}, as well as all the properties calculated for these structures in Paper 1 as described in Section~5 and published in the tables of that paper's Appendix~A. These properties include the mass ($M$), radius ($R$), velocity dispersion ($\sigma_v$), virial parameter (\alphavir), surface density ($\Sigma$), external pressure ($P_e$), and free-fall time ($t_{ff}$). 

Dendrograms are hierarchical and include substructure at many size scales, so are preferred for analyses where including the various size scales of the molecular clouds is useful, such as in a size-linewidth plot. ``Leaves'' are the smallest of the dendrogram structures, having no substructure of their own. Leaves are bounded by structures called ``branches'' and ``trunks'', where trunks are the largest and are not bounded by any further structures.  However, because of this hierarchical nature, dendrograms count emission {multiple times}, and so cannot be used in any sort of counting statistic, such as examining distributions of properties in a histogram. In these cases, we use clumps instead, which have no overlap and so do not multiply count emission. We use dendrogram structures in Sections\,\ref{sec:SL plots} and \ref{sec: virial plot} and we use clumps in Section\,\ref{sec:property comparison}.

\subsection{Region Selection} \label{subsec:regions}

\begin{figure}
    \centering
    \includegraphics[width=0.45\textwidth]{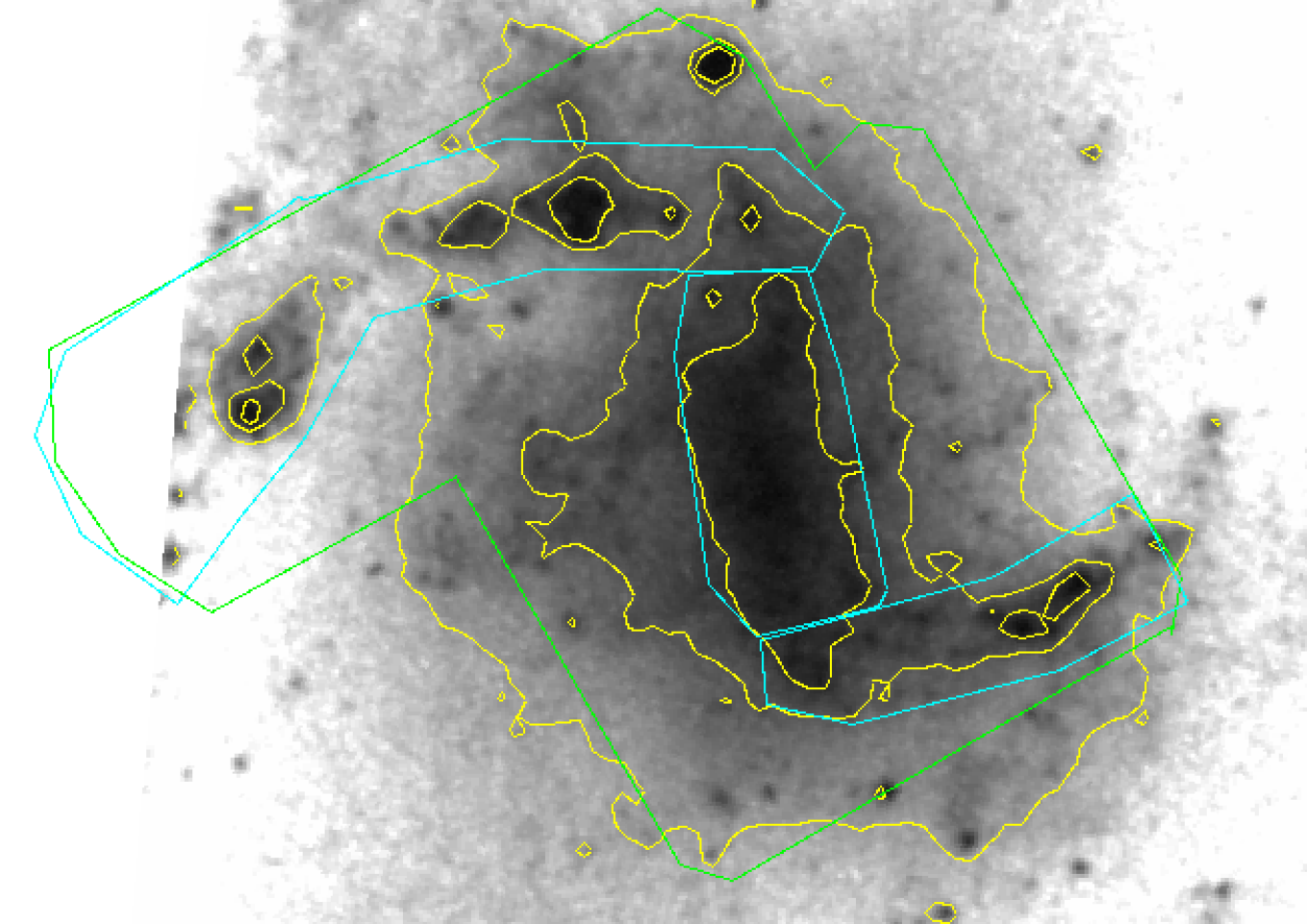}
    \includegraphics[width=0.45\textwidth]{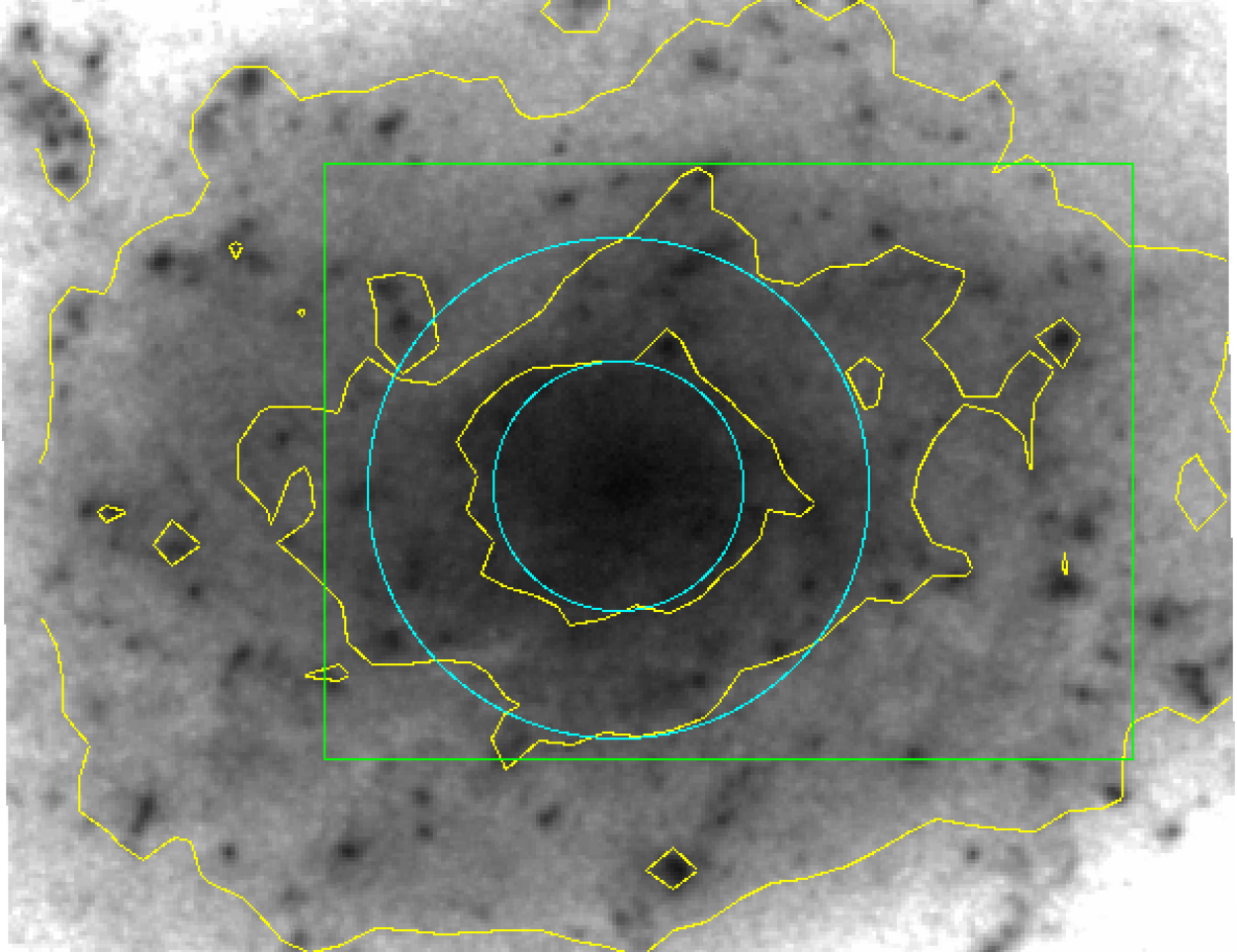}
    \caption{DSS2-red images of NGC~1313 (top) and NGC~7793 (bottom) in grayscale, with the yellow contours tracing 40\%, 60\%, and 75\% of the maximum brightness in each image. The green outlines show the observational footprints of the ALMA CO(2-1) maps, and the defined regions are shown in cyan. In NGC~1313, those regions are the northern arm, bar, and southern arm, and in NGC~7793 they are the ``center'' and ``ring'' regions. The clouds that are not inside any of the regions are defined to belong to the interarm region of NGC~1313 and the outer region of NGC~7793.}
    \label{fig:galactic regions}
\end{figure}

{In this work, we are interested in how the properties of the two galaxies vary by region within the galaxies.} To define these regions, we use contours of the red filter from Digitized Sky Survey (DSS) images of the two galaxies, which traces the bulk of the stellar population. These contours and the defined regions are shown in Figure\,\ref{fig:galactic regions}.

For NGC~1313, we define regions for the bar, the northern arm, and the southern arm, and any structures that fall outside of those regions are considered interarm (Figure\,\ref{fig:galactic regions}). The bar region is based on the 75\% brightness contour in the DSS2-red image, while the arm regions enclose the emission within the 60\% contours, then follow the arm pattern out to the edges of the CO(2-1) map following the 40\% contour. We separately consider the northern and southern arms.

Since NGC~7793 is a flocculent spiral and does not have clearly defined arms, we instead split the galaxy up into a circular ``center'' region that follows the 75\% brightness contour in the DSS2-red image, and then a ``ring'' region surrounding the center following the 60\% contour (Figure\,\ref{fig:galactic regions}). The rest of the structures are considered part of the ``outer'' region.

\begin{table*}
    \centering
    \caption{Number of cloud structures and clusters in each galaxy and region}
    \begin{tabular}{ll | ccccc | cccc}
        \hline
        \hline
        \rowcolor{white}&&\multicolumn{5}{c}{Molecular Gas Structures} & \multicolumn{4}{c}{LEGUS Clusters} \\
         Galaxy & Region & Trunks & Branches & Leaves & Clumps & Massive Clumps$^a$ & All & Young$^b$ & Massive$^a$ & YMCs$^a,b$ \\
         \hline
         NGC 1313 & All  & 65 & 82 & 442 & 531 & 137 & 1201 & 618 & 333 & 37 \\
          & Bar          & 7  & 1  & 61  & 69  & 7   & 380  & 184 & 107 & 8 \\
          & N Arm        & 20 & 32 & 161 & 193 & 62  & 297  & 243 & 48  & 18 \\
          & S Arm        & 27 & 44 & 156 & 188 & 55  & 145  & 93  & 32 & 7 \\
          & Interarm     & 11  & 5  & 64  & 82 & 13  & 379  & 98  & 146 & 4 \\
          \hline
         NGC 7793 & All & 130 & 187 & 761 & 965 & 306 & 467 & 296 & 53 & 3\\
          & Center      & 22  & 63  & 162 & 203 & 81  & 57  & 22  & 7  & 0 \\
          & Ring        & 46  & 63  & 277 & 354 & 118 & 108 & 38  & 15 & 2 \\
          & Outer       & 62  & 61  & 322 & 408 & 107 & 302 & 109 & 31 & 1 \\
         \hline
    \end{tabular} \\
    {Note:} $^a$The threshold used to define ``massive'' for both clumps and clusters in this table is $M>10^4 M_\odot$. \\ $^b$ The threshold used to define ``young'' for clusters is an age~$<10$~Myr.
    \label{tab:segmentation}
\end{table*}

\section{Cluster Catalogs} \label{sec: cluster catalogs}

{As in Paper 1, we use the catalogs of identified star clusters and their SED-fitted masses, ages, and extinctions from the LEGUS collaboration, which follow the methodology of \cite{Adamo17}. For NGC~1313 and NGC~7793, we expect the 90\% mass completeness limit of these catalogs to be approximately 1000 M$_\odot$ for clusters up to ages of 200 Myr. Further details about the cluster catalogs used in this study can be found in Paper 1. }

{In Table\,\ref{tab:segmentation}, we report the number of clusters in each galaxy as well as each of the sub-galactic regions in each galaxy. Also shown are the number of clusters that are massive (>$10^4$ M$_\odot$), young (<10 Myr), or both. We also report in this table the numbers of molecular gas structures, and how many clumps are massive (>$10^4$ M$_\odot$). This table is adapted from Table~3 in Paper 1, which gave these numbers for the galaxies as a whole. }

From Table\,\ref{tab:segmentation} we see that the molecular clouds and star clusters in NGC~7793 are fairly evenly distributed (since the area of the center, ring, and outer regions are consecutively larger), with only the exception of a slight excess of massive gas clumps in the ring region. In NGC~1313, the molecular clouds are much more abundant in the arms than in the bar or interarm regions, but there are more clusters in the bar and interarm regions than in the arms, especially for massive clusters. However, the region with the most young clusters and young, massive clusters is the northern arm. A similar excess is not seen in the southern arm. This is surprising since we expect the two spiral arms to behave in similar ways, {though perhaps not so surprising given that} the southern arm is closer to the site of a burst in star formation \citep{Larsen07, SilvaVillaLarsen12}.

\subsection{Cluster Property Distributions}

\begin{figure*}
    \centering
    \includegraphics[width=0.45\textwidth]{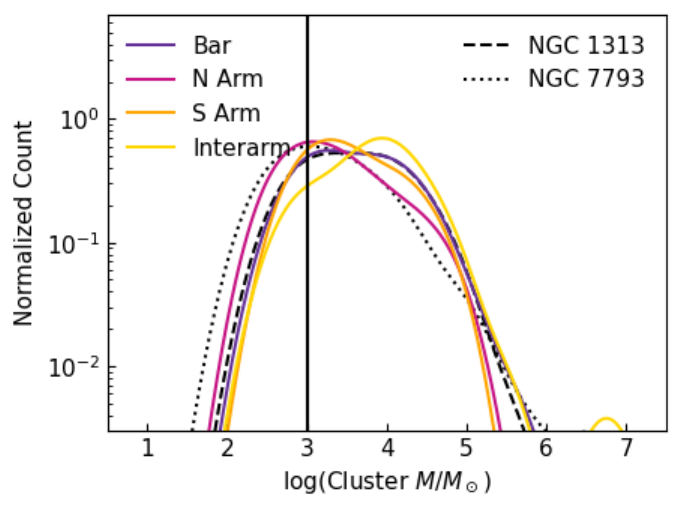}
    \includegraphics[width=0.43\textwidth]{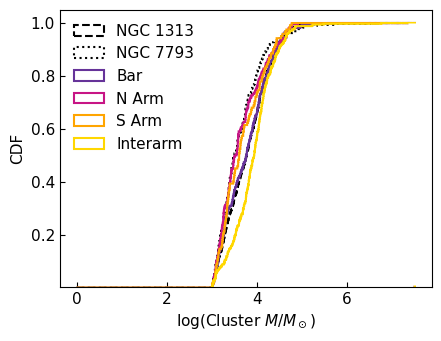}
    \includegraphics[width=0.45\textwidth]{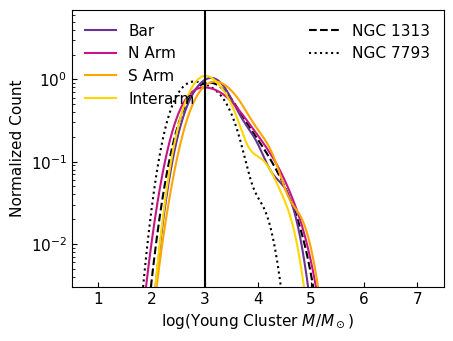}
    \includegraphics[width=0.43\textwidth]{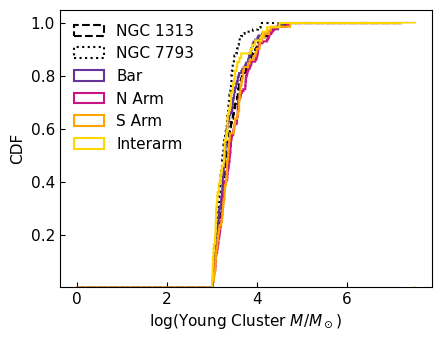}
    \includegraphics[width=0.45\textwidth]{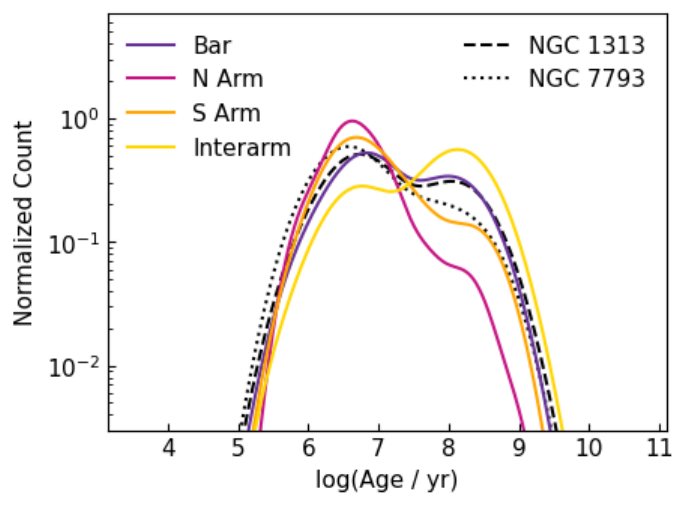}
    \includegraphics[width=0.43\textwidth]{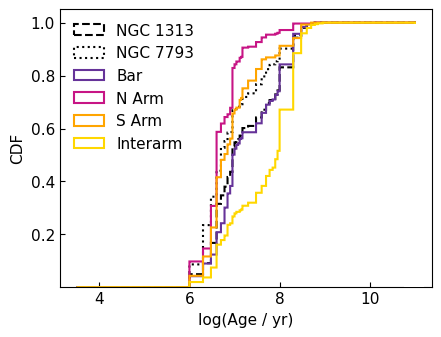}
    \caption{Distributions of the cluster parameters for the bar, northern arm, southern arm, and interarm regions of NGC~1313 using KDEs (left) and CDFs (right). The global property distributions of clusters in NGC~1313 and NGC~7793 are also shown as black dashed and dotted lines, respectively, and are the same as those in {Figure~3 of Paper 1. We show mass distributions for both the full cluster population (top) and only the young (<10 Myr) clusters (middle).} The estimated mass completeness limit of 1000~M$_\odot$ is shown as a vertical line in the top two left panels, and the CDFs of the mass distributions only includes clusters above this mass limit. The age distributions do not include clusters that are likely to have incorrect ages due to the age/reddening degeneracy \citep{Whitmore23}. }
    \label{fig:1313 cluster distributions}
\end{figure*}

\begin{figure*}
    \centering
    \includegraphics[width=0.45\textwidth]{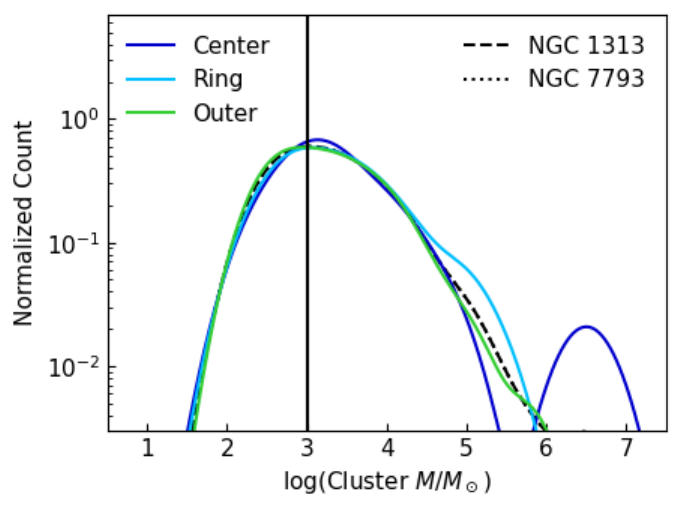}
    \includegraphics[width=0.43\textwidth]{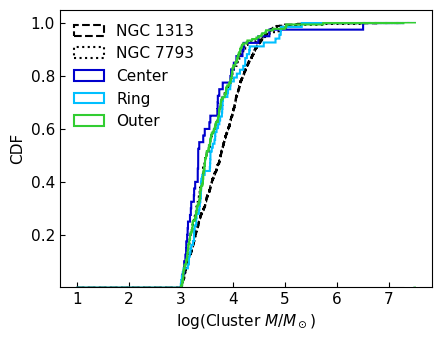}
    \includegraphics[width=0.45\textwidth]{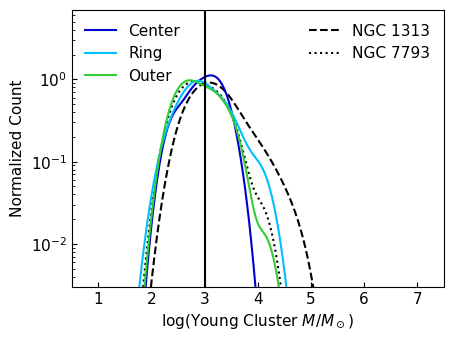}
    \includegraphics[width=0.45\textwidth]{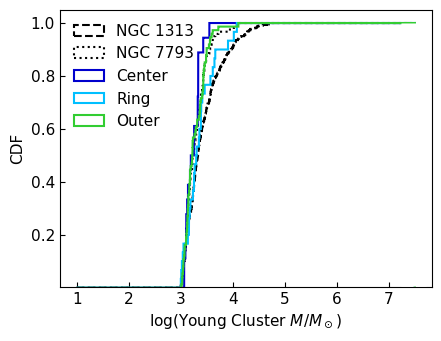}
    \includegraphics[width=0.45\textwidth]{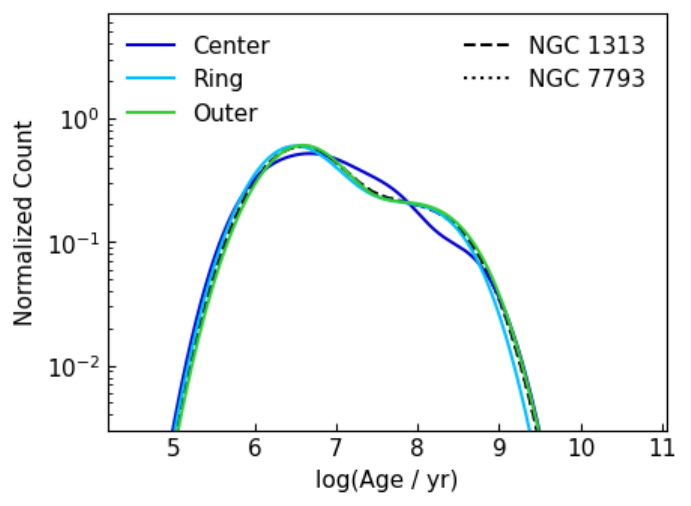}
    \includegraphics[width=0.43\textwidth]{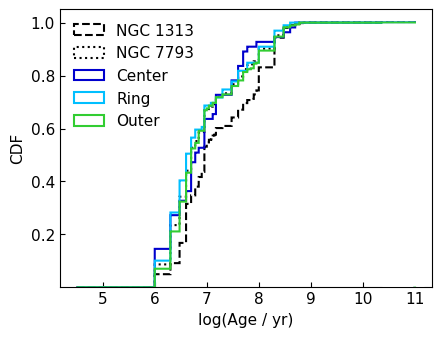}
    \caption{Distributions of the cluster parameters for the center, ring, and outer regions of NGC~7793 using KDEs (left) and CDFs (right). The global property distributions of clusters in NGC~1313 and NGC~7793 are also shown as black dashed and dotted lines, respectively, and are the same as those in {Figure~3 of Paper 1. We show mass distributions for both the full cluster population (top) and only the young (<10 Myr) clusters (middle).}  The estimated mass completeness limit of 1000~M$_\odot$ is shown as a vertical line in the top two left panels, and the CDF of the mass distribution only includes clusters above this mass limit. The age distributions do not include clusters that are likely to have incorrect ages due to the age/reddening degeneracy \citep{Whitmore23}. }
    \label{fig:7793 cluster distributions}
\end{figure*}

{To further examine how the cluster populations in each sub-galactic region compare, we plot their mass and age distributions using Gaussian kernel density estimations (KDEs) from \texttt{scipy} \citep{scipy}, and cumulative distribution functions (CDFs).  We use both of these depictions since KDEs are more intuitive to interpret, but are subject to binning effects, while CDFs do not rely on binning and so are more robust but less intuitive in understanding the underlying distribution. For figure clarity, we show only the KDEs and not the underlying histograms. For the KDEs in this section and in Section\,\ref{sec:property comparison}, we adopt a scalar estimator bandwidth of 0.5 dex to make comparisons between regions and properties as direct as possible. These distributions are shown in Figures\,\ref{fig:1313 cluster distributions} and \ref{fig:7793 cluster distributions}, and include mass distributions for both the full cluster population as well as for only the young (<10 Myr) clusters.}

In NGC~1313, there are clear, distinct differences in the mass and age distributions of {the full cluster populations} in these regions. The northern arm has much lower cluster masses than the other regions, but also is most heavily skewed towards young clusters. Meanwhile, the interarm region is more dominated by high-mass, older clusters. The southern arm and the bar have a fairly similar mass distribution, though the bar tends to have older clusters than the southern arm. When considering only the young clusters, the differences in the mass distributions between regions is much smaller {be eye}, to the point of {appearing} negligible. 

In NGC~7793, the center region has a bump at high masses ($\sim10^{6.5}$~M$_\odot$), but the CDF appears to show that all three regions have nearly identical mass distributions. These two supermassive clusters in the center of NGC~7793 are relatively old, with fitted ages of 0.9 and 12 Gyr. {We consequently we deem them largely irrelevant to this current study and will disregard them for the remainder of this work}. {When considering only the young clusters, the center region instead has the steepest mass distribution, while the outer region has the most clusters at high masses.} The age distributions between the regions are also quite similar. Overall, this suggests that the properties of the cluster population in NGC~7793 are fairly uniform throughout the galaxy.

\section{Size-Linewidth Relations} \label{sec:SL plots}

{To compare the molecular cloud properties in the regions of these two galaxies, we first look at their size-linewidth relations \citep{Larson81}. We fit power laws to the dendrogram structures in each region of the form}

\begin{equation}
    \sigma_v = a_0 R^{a_1}.
\end{equation}

{The fitted slope ($a_1$) of this relation has been measured in many environments and can vary largely based on the resolution, the molecular tracer, or the methods used to measure sizes and linewidths \citep[e.g.][]{Solomon87,Rice16,MivilleDeschenes17,Faesi16,Bolatto08,MAGMA,Nayak16,Indebetouw20,Wong22, Finn22}. These slopes are also often poorly constrained and difficult to compare. We therefore focus primarily only on comparisons between regions in this study only, and we fit only the intercept of this power law relation, holding the slope constant at a value of $a_1=0.5$. This fitted intercept indicates the relative amount of kinetic energy in clouds of a given size scale between regions. The assumed value of the fixed slope affects the value of the fitted intercept, but not the conclusions about the relative kinetic energy in the regions.}

\begin{figure}
    \centering
    \includegraphics[width=0.45\textwidth]{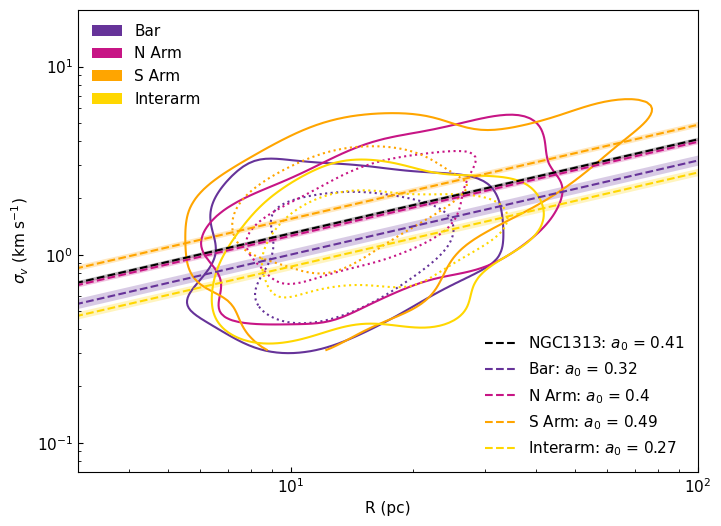}
    \includegraphics[width=0.45\textwidth]{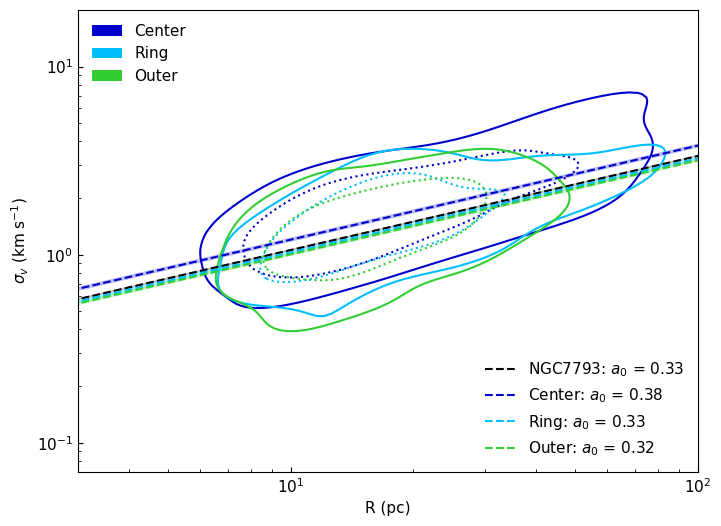}
    \caption{Deconvolved velocity dispersions plotted against deconvolved radii of dendrogram structures in the different regions of NGC~1313 (top) and NGC~7793 (bottom) showing the 20\% KDE contours as solid lines and then 50\% contours as dotted lines. The left two plots show the fully fitted power laws, with the resulting slopes shown in the lower right corners, while the plots on the right show fitted power laws with the slope held constant at $a_1=0.5$ with the resulting intercepts shown in the lower right corners. Their respective $1\sigma$ errors are shown as shaded regions. }
    \label{fig:SL regions}
\end{figure}

{In Paper 1, we fit intercepts of $a_0=0.41\pm0.01$ in NGC~1313, and $a_0=0.33\pm0.01$ for NGC~7793, indicating that the molecular clouds in NGC~1313 have significantly higher kinetic energies than those in NGC~7793.} 
The size-linewidth relations for each of the regions in the two galaxies are shown in Figure\,\ref{fig:SL regions} and the fitted values for each region and the galaxies as a whole are reported in Table\,\ref{tab:size linewidth fits}.

The regions of NGC~1313 show a spread of kinetic energies, with the southern arm having the highest intercept, followed by the northern arm, the bar, then the interarm regions. The fitted intercept for the southern arm is significantly higher than the intercepts for the bar and the interarm region, though is just barely consistent within $3\sigma$ with the northern arm. Similarly, the northern arm is significantly higher than the interarm region but consistent within $3\sigma$ with the bar. The bar and interarm regions are consistent with each other as well. These fitted intercepts suggest that the spiral arms of NGC~1313 have higher kinetic energy than the gas in the bar and interarm regions. 

In NGC~7793, the intercepts have a much smaller range than in NGC~1313, and all are consistent with each other within $3\sigma$. The center of NGC~7793 is most similar in intercept to the northern arm of NGC~1313, while the outer region is most similar to the bar. 

\begin{table}
    \centering
    \caption{Fitted intercepts for size-linewidth relations with fixed slope of $a_1=0.5$}
    \begin{tabular}{ll | c}
        \hline
        \hline
         Galaxy & Region  & Intercept, $a_0$  \\
         \hline
         NGC 1313  
          & Global    & 0.41$\pm$0.01 \\
          & Bar       & 0.32$\pm$0.02 \\ 
          & N Arm     & 0.40$\pm$0.01 \\
          & S Arm     & 0.49$\pm$0.02 \\
          & Interarm  & 0.27$\pm$0.01 \\ 
          \hline
         NGC 7793 
          & Global & 0.33$\pm$0.01  \\ 
          & Center & 0.38$\pm$0.01 \\ 
          & Ring   & 0.33$\pm$0.01 \\ 
          & Outer  & 0.32$\pm$0.01 \\
         \hline
    \end{tabular} 
    \label{tab:size linewidth fits} 
    \\
    {Note:} Global fitted intercepts from Paper 1.
\end{table}

\section{Virialization} \label{sec: virial plot} 

\begin{figure}
    \centering
    \includegraphics[width=0.45\textwidth]{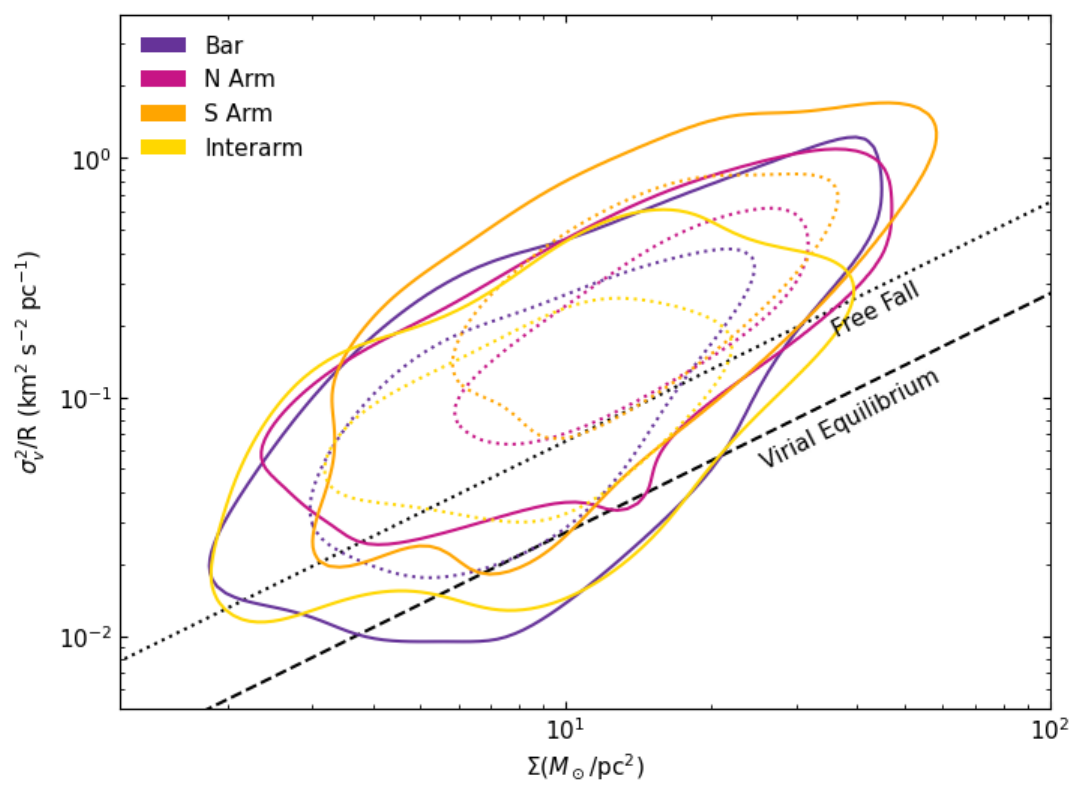}
    \includegraphics[width=0.45\textwidth]{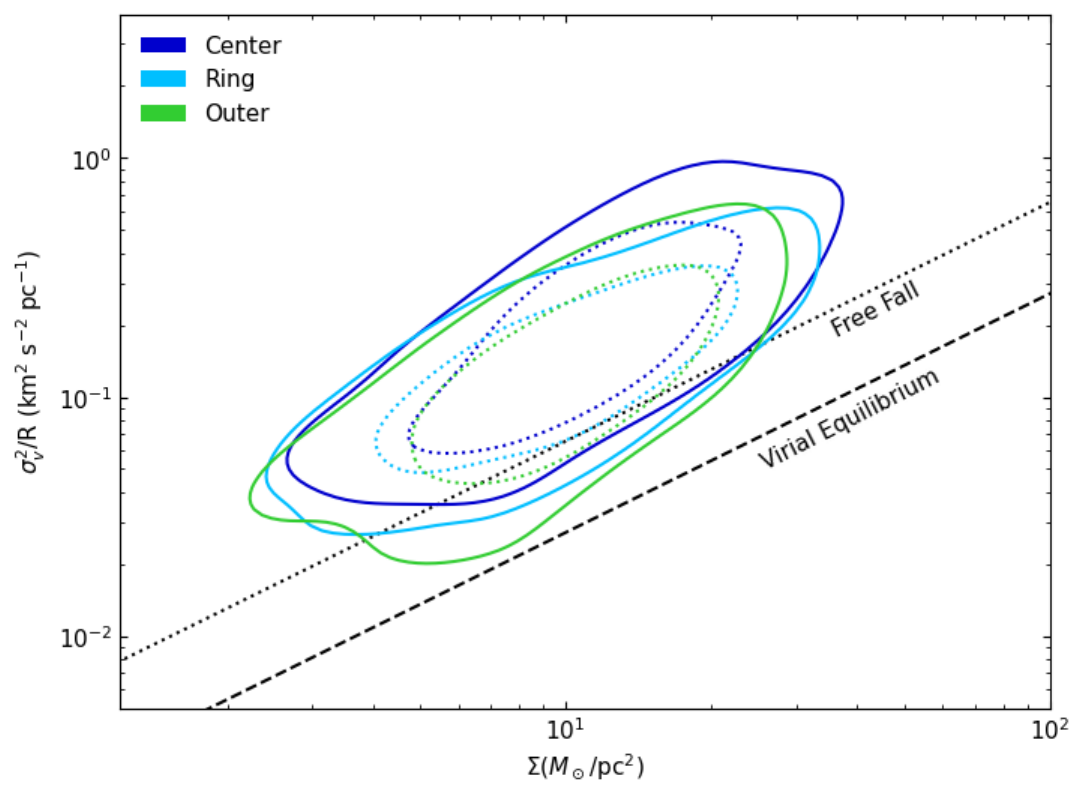}
    \caption{The velocity metric plotted against surface density of the dendrogram structures in the different regions of NGC~1313 (top) and NGC~7793 (bottom) showing the 20\% KDE contours as solid lines and the 50\% contours as dotted lines. The dashed line shows where clouds in virial equilibrium would fall, above the line being dominated by kinetic energy. Clouds that have begun free-fall collapse would also have enhanced kinetic energy and fall along the dotted line. }
    \label{fig:virial regions}
\end{figure}

{We next compare the balance between gravitational and kinetic energy in the clouds of the different regions by plotting their velocity metrics, $\sigma_v^2/R$, against their surface densities, $\Sigma$ (Figure\,\ref{fig:virial regions}). The line of virial equilibrium is shown as a dashed line, where gravity is balanced by kinetic energy. Above this line, clouds would be super-virial and dominated by kinetic energy, and below this line clouds are sub-virial and likely to collapse due to gravity dominating. Clouds can appear super-virial for many reasons, including the possibility that they are unbound and will disperse, that they are bounded by an external pressure, or that they are in free-fall, in which case they would fall along the dotted free-fall line where \alphavir=2. In Figure\,\ref{fig:virial regions}, the distribution of clouds in each region are depicted with contours of the two-dimensional KDEs showing 20\% and 50\% of the maximum density.}

In NGC~1313, all of the regions show a dip towards the virial equilibrium line, though the biggest populations close to virial equilibrium are in the bar and the interarm regions, where the 50\% density contours also show that dip. The southern arm appears to extend higher in the plot, suggesting that more of its clouds are either unbound or would require an external pressure to remain bound compared to the other regions. It seems likely that this is related to the interaction history of NGC~1313 in the southwest since simulations show interaction causing clouds to become unbound \citep{Pettitt18, Nguyen18}. Interestingly, the bar of NGC~1313 does not have more clouds in the unbound or pressure-bound region of the plot than other regions despite {containing the galactic center of NGC~1313 and the fact} that many other galactic centers exhibit high external pressures \citep{JDM13, Colombo14, Kauffmann17, Walker18, Sun18PHANGS}. {This is especially pronounced in the findings of \cite{Sun20} that velocity dispersions and external pressures are most enhanced in the centers of barred galaxies. }

In NGC~7793, the three regions once again appear more uniform than in NGC~1313, although the center region does extend higher in the plots, suggesting that it has more clouds that are unbound or require external pressure than the ring and outer regions as we would expect for a galactic center. The ring and outer regions appear very similar, suggesting there is little variation in the virialization of clouds outside of the galaxy center. None of the regions of NGC~7793 however show a dip towards virial equilibrium like that seen in every region of NGC~1313, except for slightly in the ring region.

\section{Property Distribution Comparisons} \label{sec:property comparison}

\subsection{NGC~1313 Region Comparisons} \label{subsec: 1313 dists}

\begin{figure*}
    \centering
    \includegraphics[width=0.42\textwidth]{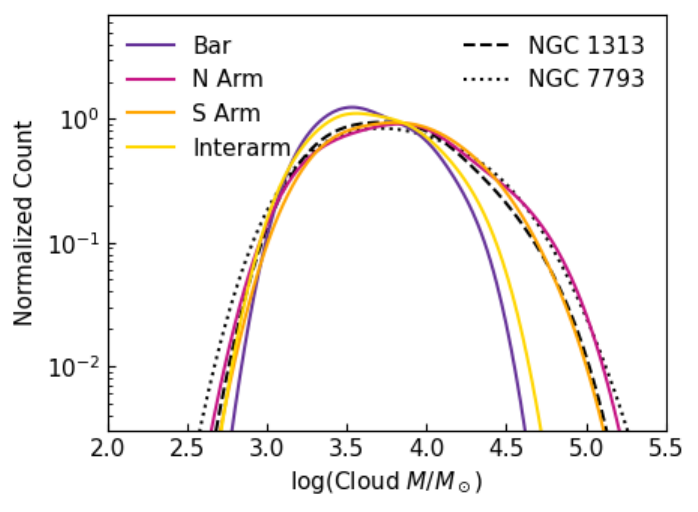}
    \includegraphics[width=0.41\textwidth]{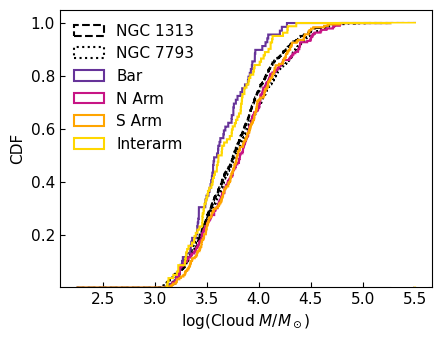}
    \includegraphics[width=0.42\textwidth]{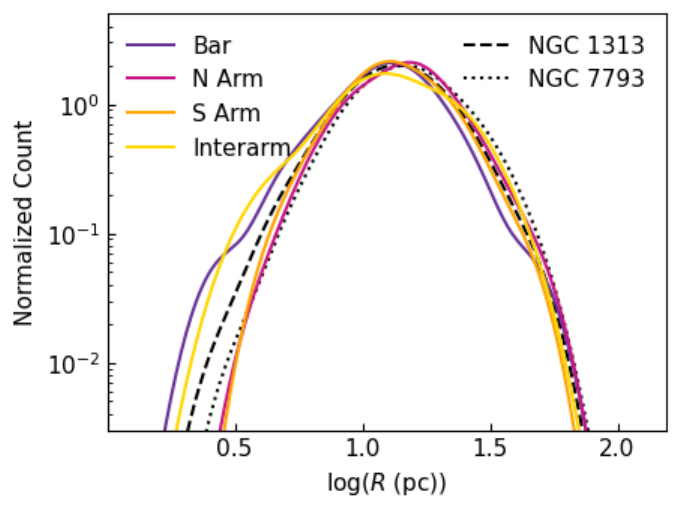}
    \includegraphics[width=0.41\textwidth]{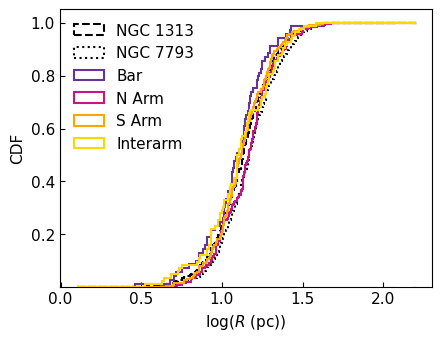}
    \includegraphics[width=0.42\textwidth]{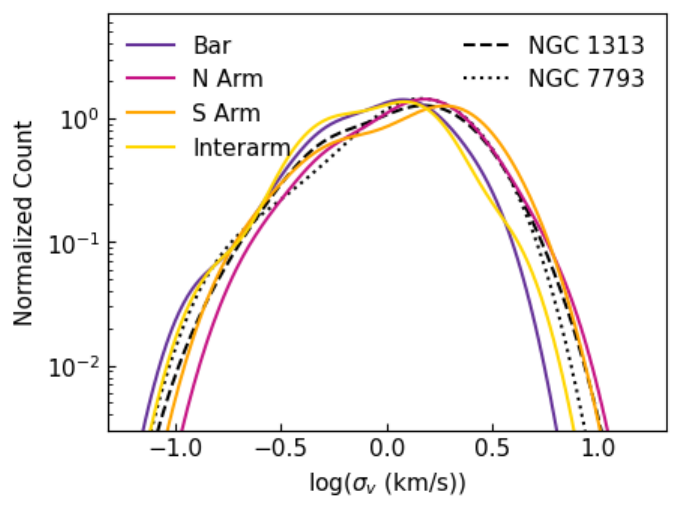}
    \includegraphics[width=0.41\textwidth]{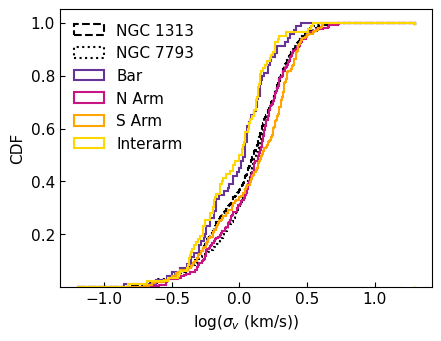}
    \caption{Distributions of the observed parameters for the bar, northern arm, southern arm, and interarm regions of NGC~1313 using KDEs (left) and CDFs (right). Also shown are the global distributions for each galaxy from Paper 1. }
    \label{fig:1313 distributions observed}
\end{figure*}

\begin{figure*}
    \centering
    \includegraphics[width=0.42\textwidth]{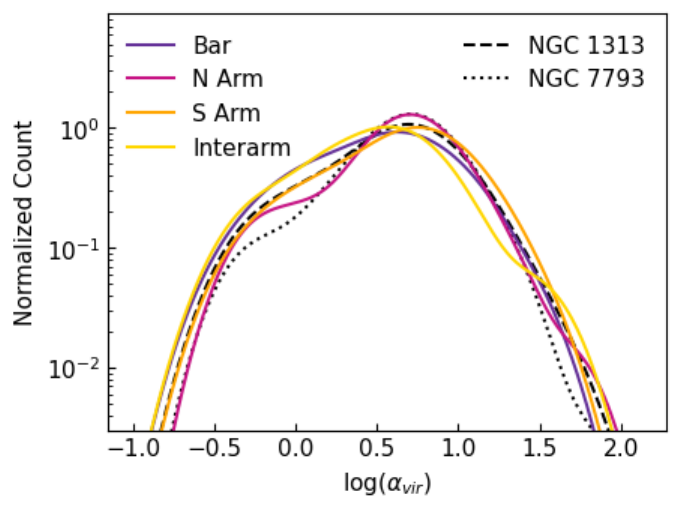}
    \includegraphics[width=0.41\textwidth]{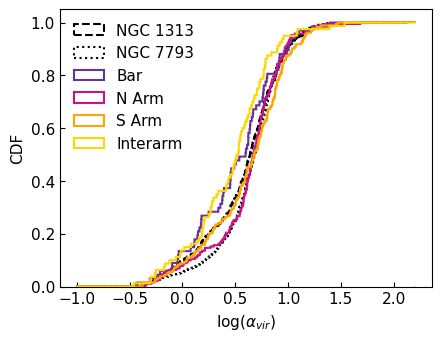}
    \includegraphics[width=0.42\textwidth]{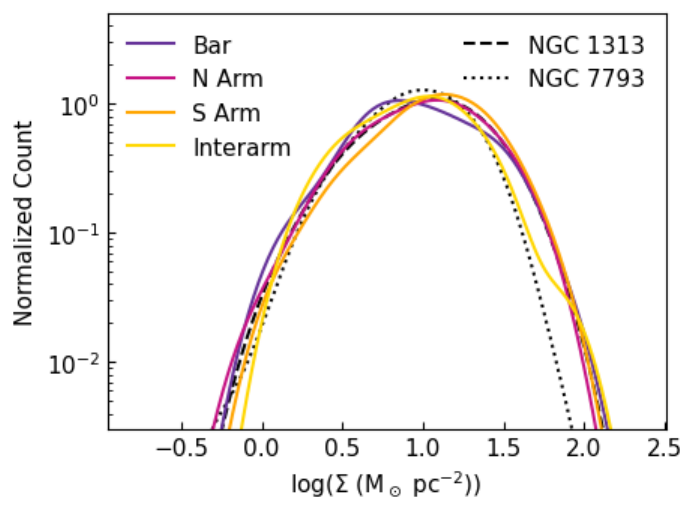}
    \includegraphics[width=0.41\textwidth]{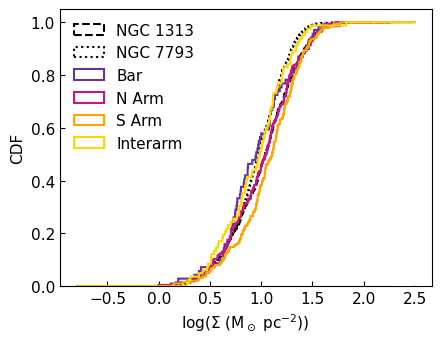}
    \includegraphics[width=0.42\textwidth]{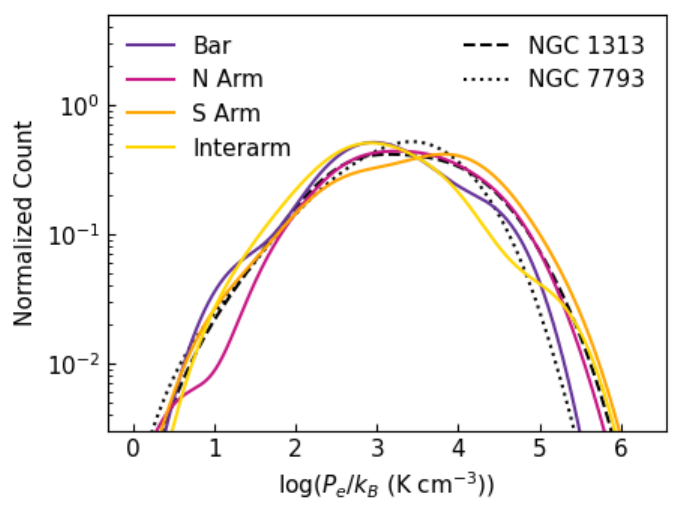}
    \includegraphics[width=0.41\textwidth]{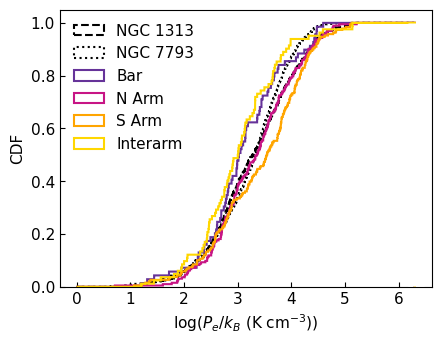}
    \includegraphics[width=0.42\textwidth]{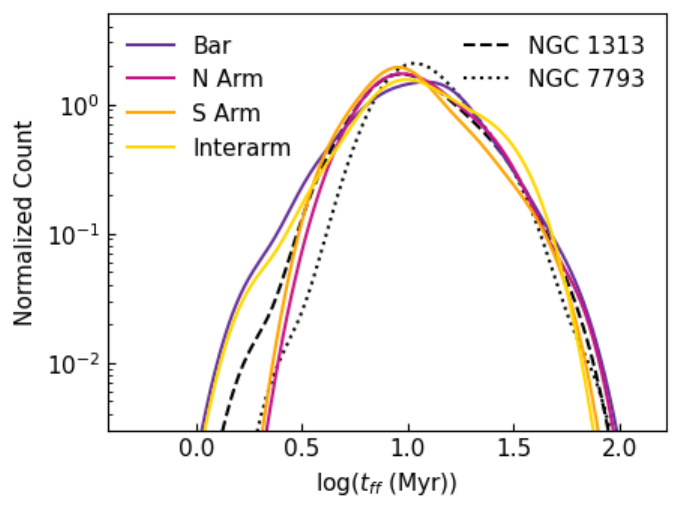}
    \includegraphics[width=0.41\textwidth]{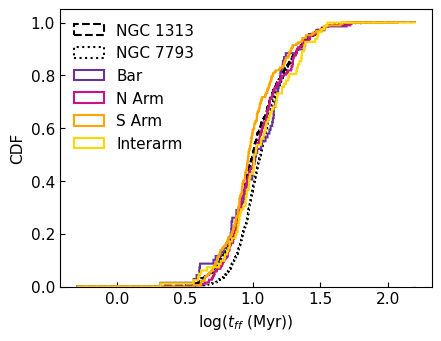}
    \caption{Distributions of the derived parameters for the bar, northern arm, southern arm, and interarm regions of NGC~1313 using KDEs (left) and CDFs (right). Also shown are the global distributions for each galaxy from Paper 1. }
    \label{fig:1313 distributions derived}
\end{figure*}

{We next compare the distribution of cloud properties within the bar, northern arm, southern arm, and interarm regions of NGC~131 by looking at KDEs and CDFs of the non-overlapping clump structures. The distributions of the global NGC~1313 and NGC~7793 cloud population from Paper 1 are shown as well for comparison.} The observed mass, radius, and linewidth distributions are shown in Figure\,\ref{fig:1313 distributions observed}, and the properties derived from them are shown in Figure\,\ref{fig:1313 distributions derived}. For figure clarity, we show only the KDEs and not the underlying histograms. 

There is a large difference in the mass distributions, with the spiral arms having significantly more massive clouds than the bar or interarm regions. The northern arm also appears to have slightly higher masses than the southern arm. These regional mass distribution differences agree with previous results that spiral arms truncate at higher masses than clouds in the interarm regions, seen in both observations \citep{Koda09, Colombo14, Rosolowsky21} and simulations \citep{Pettitt18, Nguyen18, Dobbs19}. The differences between cloud size distributions are minimal, though the bar region appears to have slightly fewer clouds with large radii. The spiral arms have similar linewidth distributions and are higher than the other two regions, where the interarm region has the smallest linewidths, again matching previous results \citep{Colombo14, Sun20, Rosolowsky21, Koda23}. 

Despite the large difference in mass distributions, the properties derived from the mass, radius, and linewidth do not show strong distinctions between regions. Most notably, the interarm and bar regions appears to have the lowest virial parameters, surface densities, and external pressures. The southern arm also appears to have slightly higher surface densities and pressures compared to the northern arm and other regions. 

It is particularly surprising that we see so little difference in surface densities between arm and interarm regions, since that has often been a notable difference in other studies \citep{Colombo14, Sun20, Rosolowsky21, Koda23}. We also note that the clouds in the bar of NGC~1313 mostly have less extreme properties than the arms, whereas other studies have often seen enhancements of the velocity dispersion, surface density, and pressure in bars and galactic centers \citep{Sun20, Rosolowsky21,Ali23}. This may not be so surprising since \cite{Querejeta21} report large variations in the properties of molecular gas in bars, potentially because star formation in bars is episodic.

{As in Paper 1, we also quantify these differences in property distributions using bootstrapped two-sample Kolmogorov-Smirnoff (KS) tests and Anderson-Darling (AD) tests. Using the full sample size of the galaxies results in an overpowered statistic as described in \cite{Lazariv18}, where the tests detect every small variation in the distribution, even those that are well below the measurement errors. Since these overpowered tests are unreliable, we instead perform a bootstrapped version, taking 1000 random subsamples and reporting the average $p$-value of the tests performed on the subsamples. We choose a sample size of 65 because we are limited by the the number of clumps identified in the bar of NGC~1313. We perform these bootstrapped tests for every property and for every combination of sub-galactic region in both galaxies and plot the results in Appendix\,\ref{append: KS AD plots}. }

{Looking at the bootstrapped KS and AD tests for pairings of the regions in NGC~1313, we find that }
none of the derived properties have statistically significant differences ($p$-values<0.05) in their distributions between regions for both KS and AD tests except for the southern arm having significantly higher pressure than the bar and interarm regions. There are, however, statistically significant differences in their mass and linewidth distributions. The absolute value of these test results are highly subject to the size of the subsamples used in the bootstrapping, and so we do not take these as indications of whether or not the distributions are actually different, but rather that some properties are {more likely to be different} than others.

\subsection{NGC~7793 Region Comparisons} \label{subsec: 7793 dists}

\begin{figure*}
    \centering
    \includegraphics[width=0.42\textwidth]{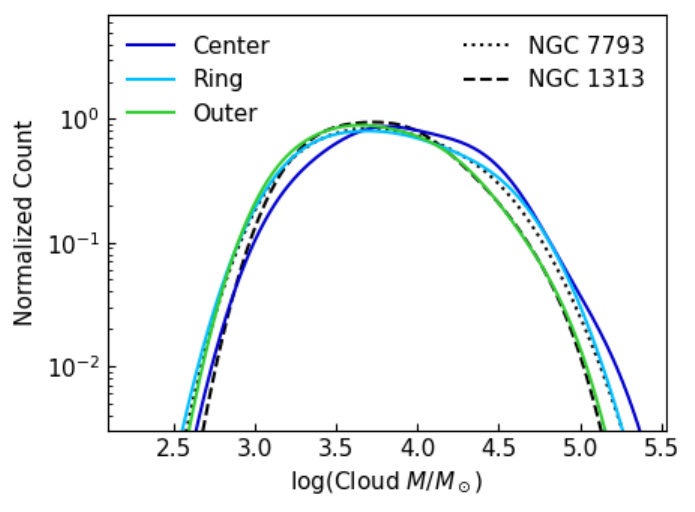}
    \includegraphics[width=0.41\textwidth]{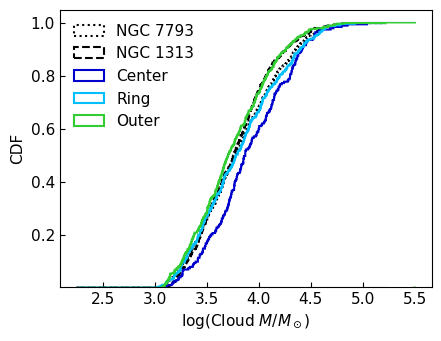}
    \includegraphics[width=0.42\textwidth]{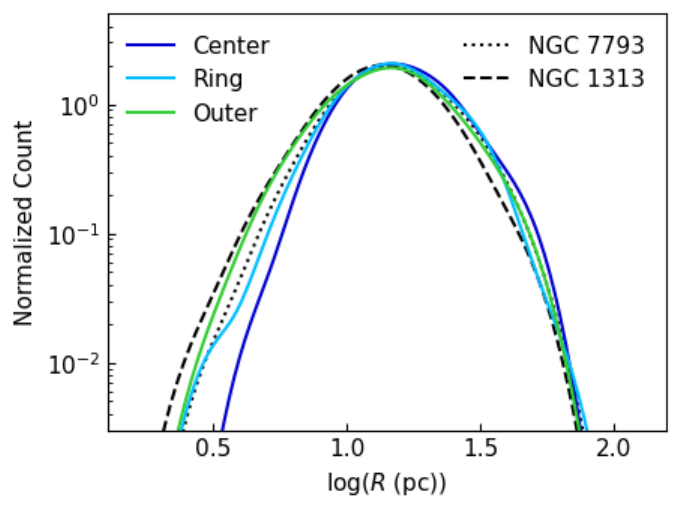}
    \includegraphics[width=0.41\textwidth]{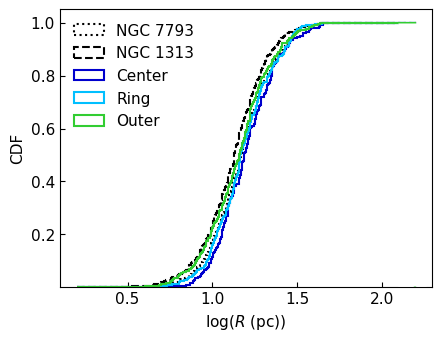}
    \includegraphics[width=0.42\textwidth]{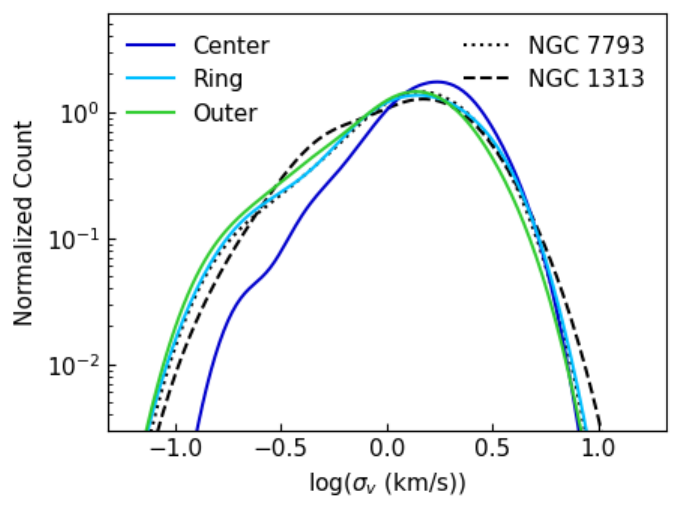}
    \includegraphics[width=0.41\textwidth]{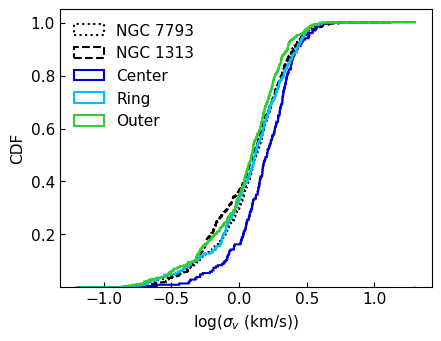}
    \caption{Distributions of the observed parameters for the center, ring, and outer regions of NGC~7793 using KDEs (left) and CDFs (right). Also shown are the global distributions for each galaxy from Paper 1.}
    \label{fig:7793 distributions observed}
\end{figure*}

\begin{figure*}
    \centering
    \includegraphics[width=0.42\textwidth]{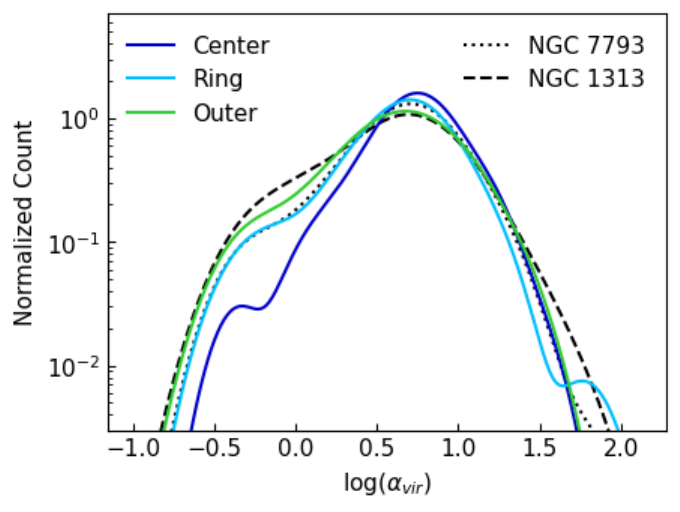}
    \includegraphics[width=0.41\textwidth]{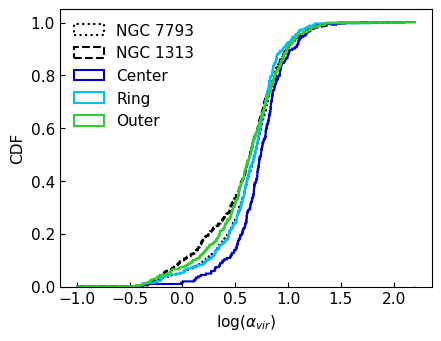}
    \includegraphics[width=0.42\textwidth]{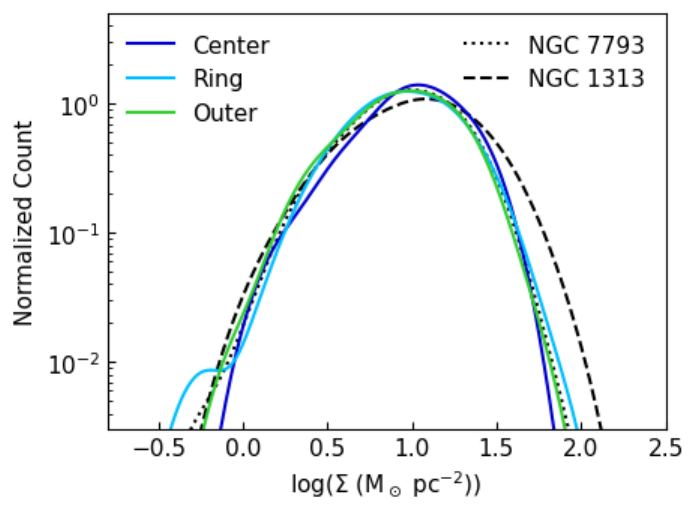}
    \includegraphics[width=0.41\textwidth]{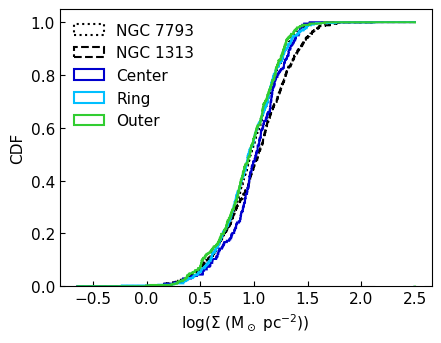}
    \includegraphics[width=0.42\textwidth]{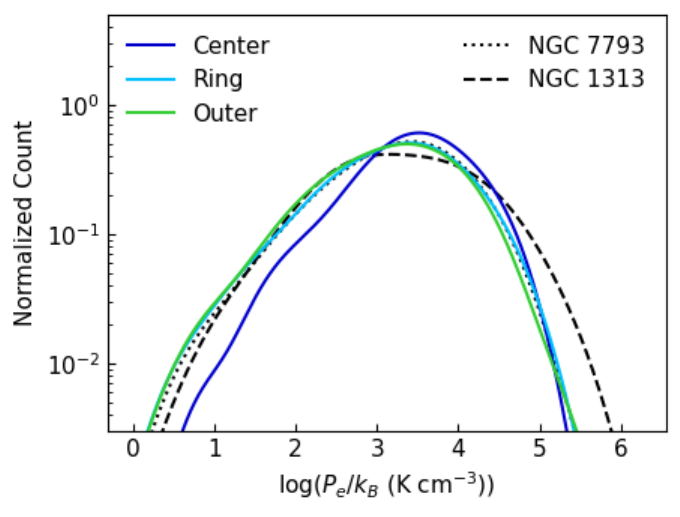}
    \includegraphics[width=0.41\textwidth]{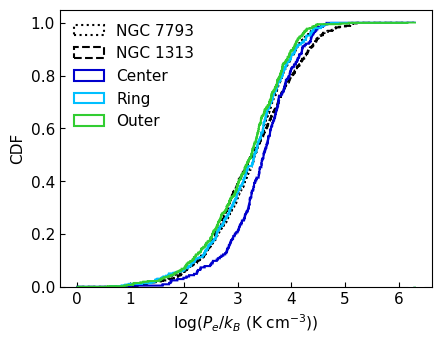}
    \includegraphics[width=0.42\textwidth]{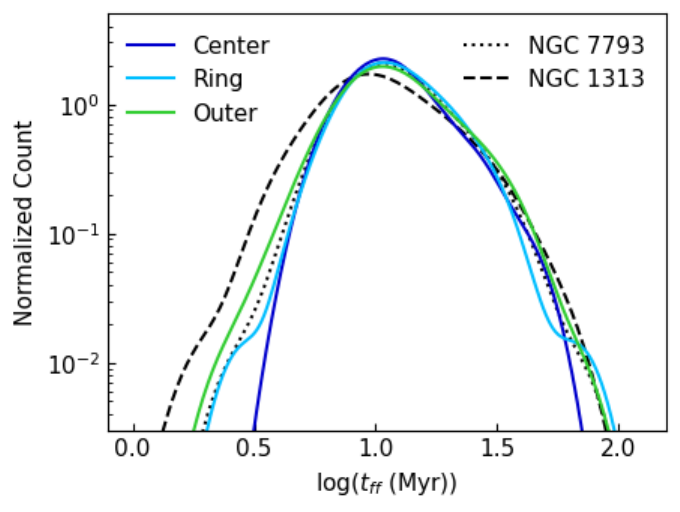}
    \includegraphics[width=0.41\textwidth]{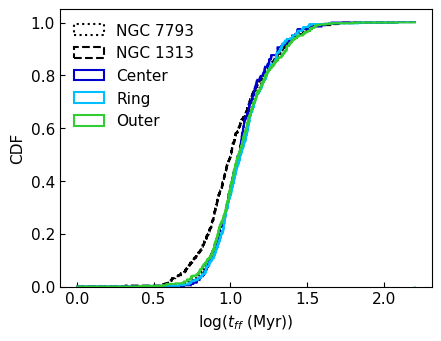}
    \caption{Distributions of the derived parameters for the center, ring, and outer regions of NGC~7793 using KDEs (left) and CDFs (right). Also shown are the global distributions for each galaxy from Paper 1.}
    \label{fig:7793 distributions derived}
\end{figure*}

We next examine the distribution of cloud properties within the center, ring, and outer regions of NGC~7793. The observed mass, radius, and linewidth distributions are shown in Figure\,\ref{fig:7793 distributions observed}, and the properties derived from them are shown in Figure\,\ref{fig:7793 distributions derived}.

The distributions in NGC~7793 vary from the center of the galaxy to the outer region, with the center having more extreme properties, such as higher masses, higher linewidths, higher virial parameters, and higher external pressures. This generally matches what was found in other surveys for the centers of galaxies \citep{JDM13,Colombo14,Sun20, Koda23}. There is not as much difference, however, in the distributions of radii, surface densities, or free-fall times. In most cases, the ring and outer regions have very similar distributions, while the center deviates more. 

These similarities and differences appear as well in comparisons of the bootstrapped KS and AD tests shown in Appendix\,\ref{append: KS AD plots}. None of the cloud or cluster properties, either observed or derived, have statistically significant differences ($p$-values<0.05) in their distributions among the regions of NGC~7793 for both KS and AD tests. The only exception is the center region having significantly higher linewidths than the outer region based on the AD test. Again, the absolute value of these test results are highly subject to the size of the subsamples used in the bootstrapping, so we do not take these results to indicate whether or not the distributions are actually different, but rather that some properties are {more likely to be different} than others.

\subsection{Inter-Galaxy Region Comparisons}

Based on the regional distributions of NGC~1313 shown in Figures\,\ref{fig:1313 distributions observed} and \ref{fig:1313 distributions derived}, none of the regions of NGC~1313 have similar property distributions to the global distributions of NGC~7793 clouds across all properties. The northern arm has the most similar mass distribution and \alphavir distribution to NGC~7793, but the interarm region has the most similar surface density distribution, and in other property distributions, such as external pressure and free-fall time, no NGC~1313 region is similar to the clouds in NGC~7793. This suggests that the environment of a flocculent spiral like NGC~7793 is quite different from both clearly defined arms and their interarm regions. Most notably, every region of NGC~1313 contains, on average, clouds with higher surface densities and shorter free-fall times than NGC~7793. 

This differences between the galaxies' regions is seen again in Figures\,\ref{fig:7793 distributions observed} and \ref{fig:7793 distributions derived}, which show that none of the regions of NGC~7793 have similar virial parameters, surface densities, external pressures, or free-fall times to the global distributions of NGC~1313. The ring region of NGC~7793 appears to have the most similar mass, radius, and linewidth distributions to the global distribution of NGC~1313.

Considering the comparisons of the bootstrapped KS and AD tests in Appendix\,\ref{append: KS AD plots}, some of the most significant differences among all the pairings for both tests are between the central region of NGC~7793 and the bar and interarm regions of NGC~1313, for almost every property, with the notable exception of surface density and free-fall time. This is somewhat surprising since the bar of NGC~1313 encompasses that galaxy's own central region {and galaxy centers usually have more extreme cloud properties \citep{JDM13,Colombo14,Sun20,Koda23,Ali23}}. One other notable difference from the KS and AD tests is that the southern arm of NGC~1313 has statistically significant lower free-fall times than all regions of NGC~7793. This is not true for the northern arm. Since the absolute value of resulting $p$-values of these bootstrapped tests are so easily influenced by subsample size, we do not take these results to indicate whether or not the distributions are actually statistically different, but rather to indicate which property distributions are {most likely to be different.}

\section{Discussion} \label{sec:discussion}

In Table\,\ref{tab:comparisons}, we briefly outline the differences in each property between the two galaxies {as discussed in Paper 1} and among the regions within each galaxy. There does not appear to be any singular property that has an out-sized influence on the star formation of either galaxy or galactic region. 

\begin{table*}[t]
    \centering
    \caption{Comparison of observed properties between environments}
    \begin{tabular}{c|c c c}
    \hline
    \hline
        Property & Global Difference & NGC~1313 Regional & NGC~7793 Regional \\
        \hline
        Cluster mass distribution & NGC~1313 more massive & Interarm most and N arm least massive & No difference \\
        Young Cluster mass distribution & NGC~1313 more massive & No difference & Outer most massive \\
        Cluster age distribution & NGC~1313 older & Interarm older, N arm youngest & No difference \\
        Size-Linewidth intercept & No difference & S arm higher, interarm lower & Center higher \\
        Virialization plot & NGC~1313 closer to virial & Bar and interarm closer to virial & Center further from virial \\
        Cloud Mass distribution & No difference & Arms more massive & Center more massive \\
        Radius distribution & NGC~7793 larger & No difference & No difference \\
        Linewidth distribution & No difference &  Arms higher & Center higher \\
        Virial parameter distribution & NGC~1313 lower & Interarm lower & Center higher \\
        Surface density distributions & NGC~1313 higher & S arm higher, interarm lowest & No difference \\
        External pressure distribution & NGC~1313 higher & S arm higher, interarm lowest & Center higher \\
        Free-fall time distribution & NGC~1313 lower & No difference & No difference \\
        \hline
    \end{tabular}
    \label{tab:comparisons}
\end{table*}

{While the differences in the cloud properties between the two galaxies as seen in Paper 1 are surprisingly small,} there is much greater variation between regions within the galaxy, especially within NGC~1313. This suggests that the local environment has a much stronger influence on cloud properties than the global galaxy environment. This was also seen when comparing regional variations and galaxy-to-galaxy variations in the PHANGS sample at 100~pc resolution \citep{Sun22}. 

Comparing the regional variations within the two galaxies, NGC~7793 appears much more uniform across the galaxy than NGC~1313, with more extreme properties only being seen in the center of the galaxy. The lack of clearly-defined arm and interarm regions makes direct comparison tricky, but it is at least clear that no {region} of NGC~7793 contains clouds with surface densities, pressures, or linewidths as high as the {most extreme properties found in the} arms of NGC~1313 {(Figure\,\ref{fig:7793 distributions observed} and \ref{fig:7793 distributions derived})}. \cite{Elmegreen09} posits that the degree of dispersion of the density probability distribution function determines the ability of a region to form massive, gravitationally bound star clusters. Regions that have a larger spread in densities will also achieve higher densities, and so be able to form massive clusters. It would follow then that NGC~1313 is able to create more massive star clusters than NGC~7793 because it has a greater variety of subgalactic environments, including strong spiral arm environments that have higher pressures and densities. {The lack of similarly large differences in the mass distributions of young clusters could be because the clusters in the LEGUS catalogs do not include many of the youngest, highly embedded clusters \citep{Messa21}, and they are no longer closely associated with their natal molecular gas, as seen in Paper 1. }

It is interesting to note as well that the southern arm of NGC~1313 has slightly more extreme properties than its northern arm. We know from measurements of HI and {the star formation history of NGC~1313} \citep{Peters94, Larsen07, SilvaVillaLarsen12, Hernandez22} that {it} is experiencing an interaction, which likely caused a recent burst in star formation in the southwest of the galaxy approximately 100 Myr ago. This interaction may also be influencing the difference in cloud properties between the northern and southern arm, suggesting that satellite galaxy interaction can drive variations in local cloud properties. 

Meanwhile in NGC~7793, its loose, poorly-defined spiral arms does not result in the majority of the clouds mimicking either the arm or the interarm regions of NGC~1313. Rather, the cloud properties throughout the galaxy have similar masses and kinetic energies to the arms of NGC~1313, but their surface densities and pressures are more similar throughout to the bar and interarm regions of NGC~1313. These differences overall result in higher virial parameters and longer free-fall times throughout NGC~7793 than any other regions in NGC~1313. Essentially, NGC~7793 has clouds that are just as massive and have just as much kinetic energy as NGC~1313, but they are puffier and less pressurized, and so are less inclined to collapse and form stars. 
This mirrors the findings in Paper 1 that the consumption time of the molecular mass in NGC~7793 is much longer than in NGC~1313. Strong spiral density waves are likely to perturb molecular clouds to induce collapse while they are in the arms, and then shear the clouds into more diffuse gas as they enter the interarm regions. The lack of strong spiral density waves in NGC~7793 allows the clouds to exist for a longer time in a dormant state where they are neither collapsing nor being torn apart. 

A notable exception to the uniformity of NGC~7793 is the center of the galaxy, where the clouds have higher masses, and fewer clouds have low linewidths, virial parameters, and pressures.  
This is reminiscent of the center of our own Galaxy, where the clouds have more extreme properties and some massive star clusters are present, but it is still forming fewer stars than expected based on the gas properties \citep{Kauffmann17,Walker18}, though this phenomenon may be {partly} accounted for by considering the metallicity dependence of the \XCO factor \citep{Evans22}.

This work represents the highest resolution direct comparison of the molecular cloud properties in spiral arm, interarm, and flocculent environments to date. Comparing our results at 13~pc to other results at $\sim$40-100~pc resolution reveals further insights about how the molecular gas behaves at different spatial scales. For the most part, we see similar trends in that spiral arms and galaxy centers have higher masses, linewidths, and pressures than interarm and outer regions of the galaxies. However, we see notably less difference in the surface densities between these regions than other studies have found at lower resolution \citep{Colombo14, Sun20, Rosolowsky21, Koda23}. This could indicate that at lower resolution, the sparse clouds of the interarm and outer regions are spread out to lower apparent surface density by the large beam size. In the arm and central regions, the clouds are sufficiently clustered that the beam is filled by clouds getting blended together, and so the apparent surface density remains high. If this is the case, the surface density seen by lower-resolution studies could be indicative of the sparsity {of} molecular clouds rather than their true {surface} density.

\section{Conclusions} \label{sec:conclusions}

{We present a comparison of the molecular gas properties in different regions of two galaxies, barred spiral NGC~1313, which is forming many massive clusters, and flocculent spiral NGC~7793, which is forming significantly fewer massive clusters despite having a similar star formation rate. Using the cloud properties calculated in Paper 1, we split the galaxies into regions including the bar, northern arm, southern arm, and interarm regions of NGC~1313 and a center, ring, and outer region of NGC~7793 since it has no clearly defined spiral arms to use. We compare how the molecular cloud properties vary by region in these two galaxies and how those regions and their differences compare between the two galaxies. Our major results are summarized below. }

\begin{itemize}

    \item The properties of the cluster population vary slightly by region, with the interarm region of NGC~1313 having the oldest and most massive clusters, while the northern arm has the youngest and least massive clusters. By number, the southern arm has significantly fewer young, massive clusters than the northern arm. The clusters in NGC~7793 show relatively little variation with region of the galaxy. However, when only the young clusters are considered, the center region of the galaxy has the fewest massive clusters.

    \item We fit power laws to the size-linewidth relations for the regions of the two galaxies, holding the slope fixed at a value of $a_1=0.5$ to determine relative kinetic energies from the fitted intercepts. The spiral arms of NGC~1313 have higher fitted intercepts, and so more kinetic energy, than the bar or interarm regions. The fitted intercepts of regions in NGC~7793 meanwhile do not differ by more than $3\sigma$. 

    \item NGC~1313 has more clouds near virial equilibrium than NGC~7793 for all regions in NGC~1313 and all regions of NGC~7793. The southern arm of NGC~1313 and the center region of NGC~7793 both show a greater spread towards the unbound region of parameter space, suggesting that more of those clouds are either unbound or would require an external pressure to remain bound.  

    \item The spiral arms of NGC~1313 tend to have more extreme cloud properties (higher masses, linewidths, surface densities, pressures, and virial parameters) than the bar or interarm regions, and they also host significantly more of the molecular gas mass. In some properties, such as linewidth, surface density, and pressure, the southern arm appears more extreme than the northern arm. This could be because the southern arm is more strongly influenced by the galaxy interaction to the southwest. The greater number of star clusters and the greater masses of those star clusters in NGC~1313 may be driven by its greater variation in environments and cloud properties within the galaxy. Its greater variation may allow for more extreme cloud properties to arise and so drive more intensive star formation. 

    \item The center region of NGC~7793 has more extreme properties than the ring and outer regions, which are quite similar to each other. This suggests that the disk of NGC~7793 is relatively uniform in cloud properties, which is consistent with finding less variation in cluster properties among the regions than in NGC~1313. The cloud properties in NGC~7793 are not particularly similar to any one region of NGC~1313, suggesting that flocculent environments are distinct from either strong spiral arms or their interarm regions. NGC~7793 has clouds that are as massive and have as much kinetic energy as NGC~1313, but have slightly larger radii and are less dense and pressurized, and so less inclined to collapse and form stars. This indicates that in NGC~7793, clouds are {likely to be dormant and form few stars for most of their lifetime}, while in NGC~1313 those clouds are perturbed by spiral density waves and either collapse and form clusters or are sheared apart into more diffuse material.

    \item We see surprisingly little variation in surface density between arm and interarm regions in NGC~1313 given previous lower-resolution results. This suggests that differences in surface densities between arm and interarm regions observed in galaxies at $\sim$40 pc or coarser resolution could be driven by variations in the number of clouds filling the beam, rather than intrinsic variations in the surface densities of the clouds themselves.

\end{itemize}

\begin{acknowledgements}

This material is based upon work supported by the National Science Foundation Graduate Research Fellowship Program under Grant No. 1842490. Any opinions, findings, and conclusions or recommendations expressed in this material are those of the author(s) and do not necessarily reflect the views of the National Science Foundation. 

KG is supported by the Australian Research Council through the Discovery Early Career Researcher Award (DECRA) Fellowship (project number DE220100766) funded by the Australian Government. 
KG is supported by the Australian Research Council Centre of Excellence for All Sky Astrophysics in 3 Dimensions (ASTRO~3D), through project number CE170100013. 
MRK acknowledges funding from the Australian Research Council through Laureate Fellowship LF220100020.

This paper makes use of the following ALMA data: ADS/JAO.ALMA\#2015.1.00782.S. ALMA is a partnership of ESO (representing its member states), NSF (USA) and NINS (Japan), together with NRC (Canada), NSC and ASIAA (Taiwan), and KASI (Republic of Korea), in cooperation with the Republic of Chile. The Joint ALMA Observatory is operated by ESO, AUI/NRAO and NAOJ. The National Radio Astronomy Observatory is a facility of the National Science Foundation operated under cooperative agreement by Associated Universities, Inc.

These data are associated with the HST GO Program 13364 (PI D. Calzetti). Support for this program was provided by NASA through grants from the Space Telescope Science Institute. Based on observations obtained with the NASA/ESA Hubble Space Telescope, at the Space Telescope Science Institute, which is operated by the Association of Universities or Research in Astronomy, Inc., under NASA contract NAS5-26555.

\facility{ALMA, HST (WFC3, ACS)}

\software{Pipeline-CASA51-P2-B v.40896 \citep{Davis21}, 
    CASA \citep[v.5.1.1-5, v.5.6.1; ][]{McMullin07}, 
    \texttt{astrodendro} \citep{Rosolowsky08}, 
    \texttt{quickclump} \citep{Sidorin17}, 
    Astropy \citep{astropy}, 
    Matplotlib \citep{matplotlib}, 
    NumPy \citep{numpy}, 
    SciPy \citep{scipy},
}

\end{acknowledgements}

\bibliographystyle{aasjournal}
\bibliography{references.bib}

\appendix

\section{Kolmogorov-Smirnoff and Anderson-Darling Tests} \label{append: KS AD plots}

As thoroughly discussed in \cite{Lazariv18}, as the sample size increases, the discerning power of KS tests increases. However, KS tests cannot take into account error in the measurements, and so it is possible for a test to become over-powered. Even a small difference in the distribution well below the measurement uncertainty can result in a rejection of the null hypothesis that the distributions are the same. This {effect} is present in AD tests as well.

To combat overpowered statistical tests, we perform a bootstrapping method to measure the difference in the distributions. {We take 1000 random subsamples with a size of 65 data points, perform KS and AD tests between each pairing of regions, and report the average $p$-value for the 1000 subsamples.} We caution that the resulting $p$-values are highly dependent on the size of the subsample used, and so these results should only be used to compare the differences between properties and regions on equally-powered statistical footing. This is meant as an indication of which pairings are the {most likely to be different}, not whether the underlying distributions are truly, statistically different. It may be possible to select a subsample size based on the error of the measurements being tested, but that is outside the scope of the current work.

\begin{figure*}
    \centering
    \includegraphics[width=0.32\textwidth]{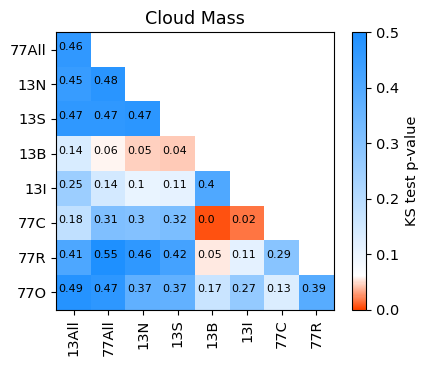}
    \includegraphics[width=0.32\textwidth]{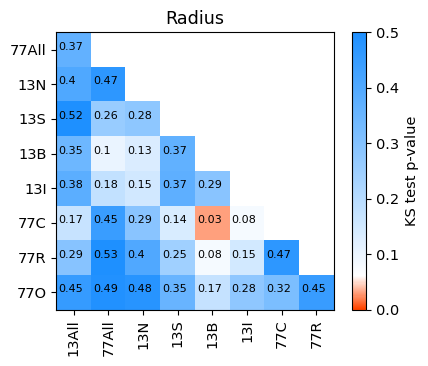}
    \includegraphics[width=0.32\textwidth]{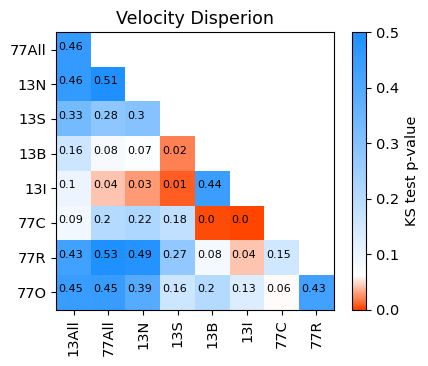}\\
    \includegraphics[width=0.32\textwidth]{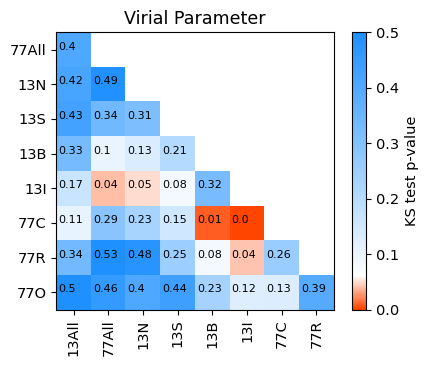}
    \includegraphics[width=0.32\textwidth]{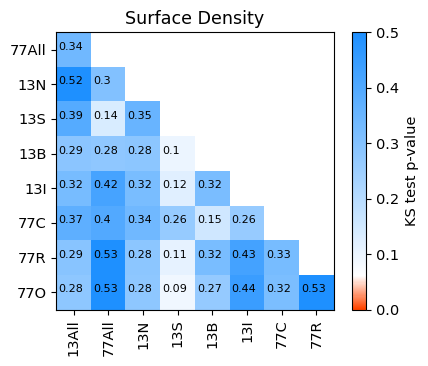}\\
    \includegraphics[width=0.32\textwidth]{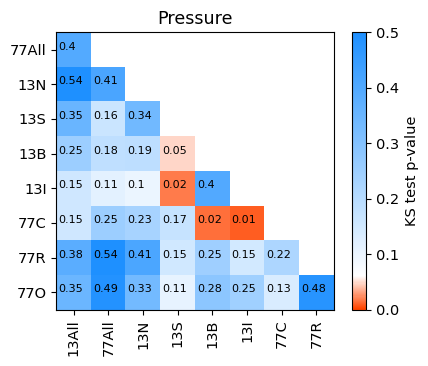}
    \includegraphics[width=0.32\textwidth]{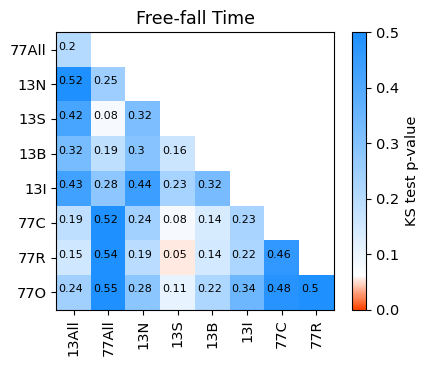}
    \includegraphics[width=0.32\textwidth]{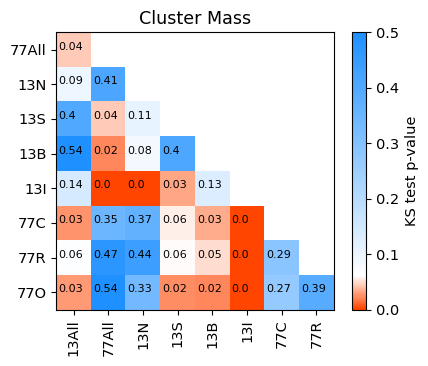}\\
    \includegraphics[width=0.32\textwidth]{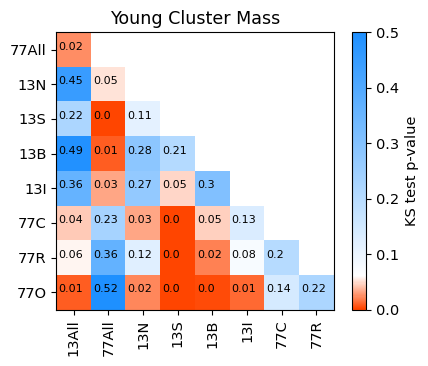}
    \includegraphics[width=0.32\textwidth]{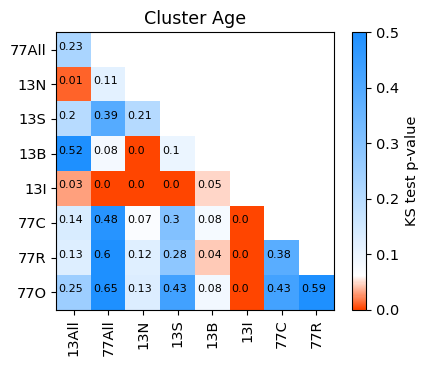}
    \caption{Bootstrapped KS tests for each cloud and cluster property and for each pairing of subgalactic region as well as global distributions for each galaxy. These $p$-values should be used as an indication of which distributions are the {most likely to be different} rather than as an absolute metric of whether any one distribution is truly different. Codes for the regions are as follows: ``13All'' is the global distribution for NGC~1313; ``77All'' is the global distribution for NGC~7793; ``13N'', ``13S'', ``13B'', and ``13I'' are the northern arm, southern arm, bar, and interarm regions of NGC~1313; and ``77C'', ``77R'', and ``77O'' are the center, ring, and outer regions of NGC~7793.}
    \label{fig:KS test grid}
\end{figure*}

\begin{figure*}
    \centering
    \includegraphics[width=0.32\textwidth]{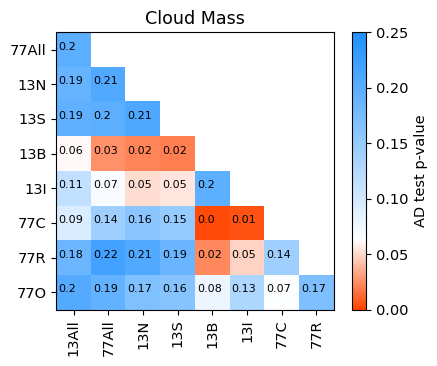}
    \includegraphics[width=0.32\textwidth]{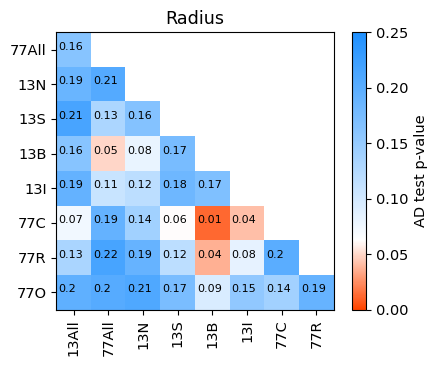}
    \includegraphics[width=0.32\textwidth]{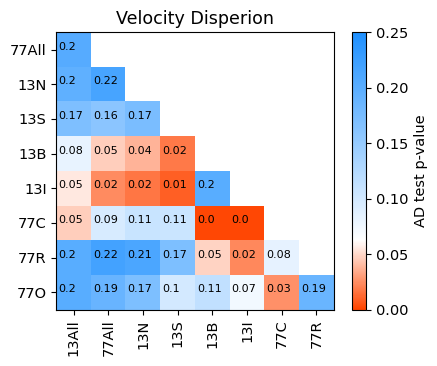}\\
    \includegraphics[width=0.32\textwidth]{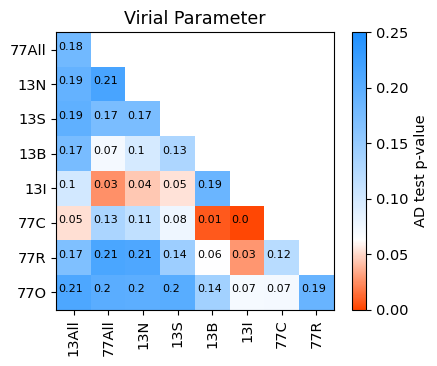}
    \includegraphics[width=0.32\textwidth]{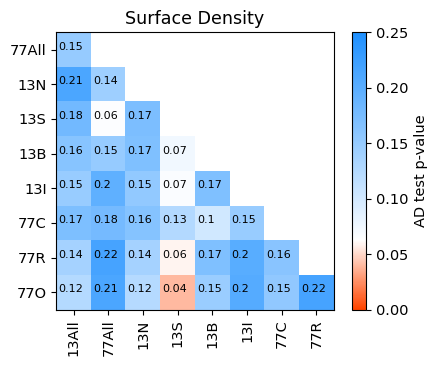}\\
    \includegraphics[width=0.32\textwidth]{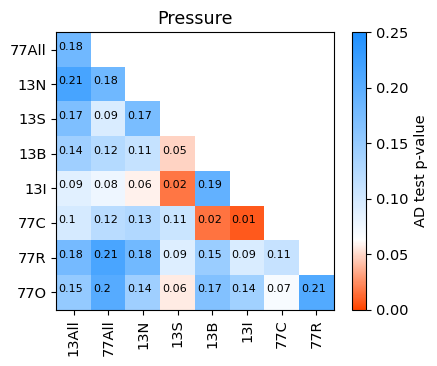}
    \includegraphics[width=0.32\textwidth]{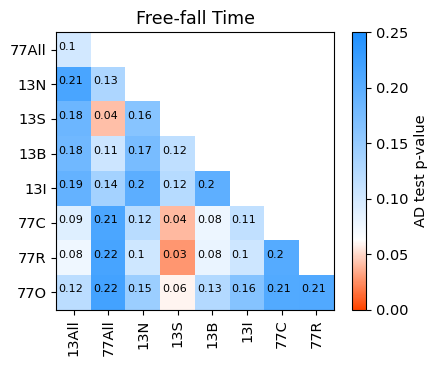}
    \includegraphics[width=0.32\textwidth]{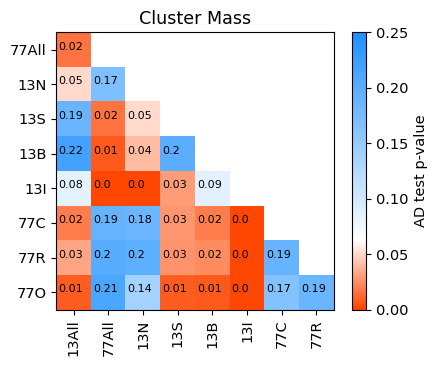}\\
    \includegraphics[width=0.32\textwidth]{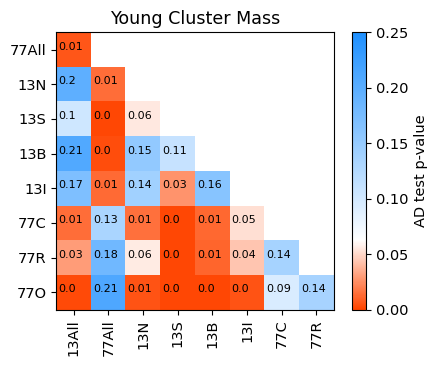}
    \includegraphics[width=0.32\textwidth]{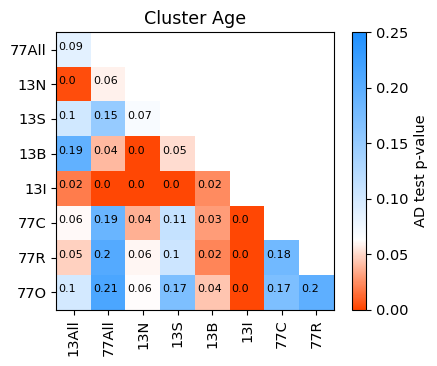}
    \caption{Bootstrapped AD tests for each cloud and cluster property and for each pairing of subgalactic region as well as global distributions for each galaxy. These $p$-values should be used as an indication of which distributions are the {most likely to be different} rather than as an absolute metric of whether any one distribution is truly different. Codes for the regions are as follows: ``13All'' is the global distribution for NGC~1313; ``77All'' is the global distribution for NGC~7793; ``13N'', ``13S'', ``13B'', and ``13I'' are the northern arm, southern arm, bar, and interarm regions of NGC~1313; and ``77C'', ``77R'', and ``77O'' are the center, ring, and outer regions of NGC~7793.}
    \label{fig:AD test grid}
\end{figure*}

Figure\,\ref{fig:KS test grid} shows the bootstrapped KS statistic for each cloud and cluster property for each pairing of subgalactic region as well as global distributions for each galaxy. Figure\,\ref{fig:AD test grid} shows the same but with an AD test. KS tests are more influenced by the center of the distribution, while AD tests are more sensitive to the tails.

\end{document}